# Dynamic variability of the phytoplankton electron requirement for carbon fixation in eastern Australian waters

David J. Hughes[1]*, Joseph R Crosswell[1,2], Martina A. Doblin[1], Kevin Oxborough[3], Peter J. Ralph[1], Deepa Varkey[1,4] and David J. Suggett[1]

[1] Climate Change Cluster, University of Technology Sydney, Broadway 2007, NSW, Australia

[2] CSIRO Oceans and Atmosphere, Brisbane, Queensland, Australia

[3] Chelsea Technologies Group Ltd, 55 Central Avenue, West Molesey, KT8 2QZ, UK

[4] Department of Molecular Sciences, Macquarie University, NSW 2109, Australia

*Corresponding author: David.Hughes@uts.edu.au

**Abstract**

Fast Repetition Rate fluorometry (FRRf) generates high-resolution measures of phytoplankton primary productivity as electron transport rates (ETRs). How ETRs scale to corresponding inorganic carbon (C) uptake rates (the so-called electron requirement for carbon fixation, $\Phi_{e,C}$), inherently describes the extent and effectiveness with which absorbed light energy drives C-fixation. However, it remains unclear whether and how $\Phi_{e,C}$ follows predictable patterns for oceanographic datasets spanning physically dynamic, and complex, environmental gradients. We utilise a unique high-throughput approach, coupling ETRs and $^{14}$C-incubations to produce a semi-continuous dataset of $\Phi_{e,C}$ ($n = 80$), predominantly from surface waters, along the Australian coast (Brisbane to the Tasman Sea), including the East Australian Current (EAC). Environmental conditions along this transect could be generally grouped into cooler, more nutrient-rich waters dominated by larger size-fractionated Chl-*a* (>10 µm) versus warmer nutrient-poorer waters dominated by smaller size-fractionated Chl-*a* (< 2 µm). Whilst $\Phi_{e,C}$ was higher for warmer water samples, environmental conditions alone explained less than 20% variance of $\Phi_{e,C}$, and changes in predominant size-fraction(s) distributions of Chl-*a* (biomass) failed to explain variance of $\Phi_{e,C}$. Instead, normalised Stern-Volmer non-photochemical quenching (NPQ$_{NSV}$ = $F_0'/F_v'$) was a better predictor of $\Phi_{e,C}$, explaining ~55% of observed variability. NPQ$_{NSV}$ is a physiological descriptor that accounts for changes in both long-term driven acclimation in non-radiative decay, and quasi-instantaneous PSII downregulation, and thus may prove a useful predictor of $\Phi_{e,C}$ across physically-dynamic regimes, provided the slope describing their relationship is predictable. We also consider recent advances in fluorescence-based corrections to evaluate the potential role of baseline fluorescence ($F_b$) in contributing to overestimation of $\Phi_{e,C}$ and the correlation between $\Phi_{e,C}$ and NPQ$_{NSV}$ – in doing so, we highlight the need for $F_b$ corrections for future field-based assessments of $\Phi_{e,C}$.

# 1. Introduction

Accurately quantifying marine primary production (MPP) at sufficient spatial and temporal scales needed to advance algorithms that retrieve carbon (C)-fixation rates from satellite ocean colour is a long-standing goal for oceanographers (Lee et al. 2015). Chlorophyll-*a* (Chl-*a*) fluorescence induction tools, such as Fast Repetition Rate fluorometry (FRRf, Kolber et al. 1998), can potentially realise this goal (e.g. Kolber & Falkowski 1993), provided FRRf-derived Photosystem II (PSII) electron transport rates per unit volume ($v\text{ETR}_{PSII}$) can be robustly converted to C-fixation rates from knowledge of the "electron requirement for carbon fixation" ($\Phi_{e,C}$, Lawrenz et al. 2013; also termed $K_C$, Hancke et al. 2015). Whilst multiple studies have demonstrated robust empirical relationships between $\text{ETR}_{PSII}$ and both gross (Suggett et al. 2009a; Robinson et al. 2014; Napoleon et al. 2013; Schuback et al. 2015), and net (Hoppe et al. 2015; Zhu et al. 2016) C-fixation, the derived values of $\Phi_{e,C}$ are highly-variable, often far exceeding the theoretical minimum stoichiometry of 4 e$^-$ (mol C)$^{-1}$ (see Kolber and Falkowski, 1993). Such variability reflects (i) re-routing of electrons to non C-fixing pathways (e.g. Fisher and Halsey, 2016), (ii) consumption of photo-produced ATP and reductant for metabolisms other than cellular growth (e.g. Halsey and Jones, 2015), (iii) growth-rate dependent variability in the lifetime of fixed-C (Halsey et al. 2011) and/or (iv) methodological bias in the determination of either $\text{ETR}_{PSII}$ or C-fixation (Suggett et al. 2009a; Hughes et al. 2018a).

Resolving variability of $\Phi_{e,C}$ over space and time poses a major challenge, since it remains unclear whether and how $\Phi_{e,C}$ follows predictable patterns, and hence, can be readily applied to broad FRRf datasets that often span complex oceanographic gradients. Through a comprehensive meta-analysis of parallel $\text{ETR}_{PSII}$ and C-fixation measurements, Lawrenz et al. (2013) demonstrated empirical relationships between $\Phi_{e,C}$ and prevailing environmental variables known to regulate photosynthesis (e.g. light, inorganic nutrients and temperature).

However, the strength of these relationships varied considerably depending upon geographic location. More recent campaigns have similarly shown that a large proportion of $\Phi_{e,C}$ variability can be explained by corresponding changes in variability of light across sites in the South China Sea (Zhu et al. 2016, 2017) or inorganic nutrients (e.g. Fe; Schuback et al. 2015, 2017; and N; Hughes et al. 2018b) over space and time. Intriguingly, several of these most recent studies revealed a role for phytoplankton community taxonomic composition in moderating the covariance between $\Phi_{e,C}$ and environmental condition. Past controlled laboratory culture experiments (Suggett et al. 2009a; Napoleon et al. 2013) and field evaluations (Suggett et al. 2006; Lawrenz et al. 2013; Robinson et al. 2014) have indicated that $\Phi_{e,C}$ appears highly variable across phytoplankton taxa. However, isolating the relative role of environment versus taxonomy upon $\Phi_{e,C}$ variability remains problematic given that certain phytoplankton functional groups are often selected-for via specific environmental conditions (see Finkel et al. 2009; Hughes et al. 2018b).

Variability of $\Phi_{e,C}$ is inherently driven by physiological 're-wiring' of the efficiency with which electrons are used to drive carbon uptake (Halsey et al. 2011; Fisher and Halsey, 2016). Thus, to overcome potential conflating roles of environment and taxonomy on regulation of $\Phi_{e,C}$, Schuback et al. (2015, 2016, 2017; Schuback and Tortell, 2019) recently considered use of a physiological trait metric as a potential overarching predictor for $\Phi_{e,C}$. These authors demonstrated an empirical relationship between the extent of non-photochemical quenching (NPQ) and $\Phi_{e,C}$, possibly driven by a positive feedback between the upregulation of non C-fixing pathways and $\Delta$pH-activation of thermal dissipation mechanisms within the PSII antennae (e.g. Nawrocki et al. 2015). This relationship appears robust for Fe-limited conditions, and also appears to hold under N-limitation (Hughes et al. 2018b), yet remains generally untested for complex coastal waters where both environmental conditions, and phytoplankton community composition are highly dynamic. The parameter

NPQ describes the dynamic downregulation of PSII photochemistry and thus captures a snapshot of the prevailing physiological status of cellular excitation energy dissipation. However, other variables such as cell size, which frequently operate as a 'master trait' capturing efficiency of resource acquisition and utilisation independently of phytoplankton taxonomic identity (Key et al. 2010; Finkel et al. 2009) also appear to show promise in broadly explaining variance in $\Phi_{e,C}$ (Zhu et al. 2017).

Currently it remains unclear whether variance of $\Phi_{e,C}$ can in fact be explained from (relatively)-easily retrieved phytoplankton properties that potentially capture both environmental and taxonomic variance across physically complex oceanographic gradients. To address this question, we employed a high-throughput coupled $ETR_{PSI}$ – $^{14}C$-incorporation technique to yield a unique, semi-continuous dataset of $\Phi_{e,C}$ ($n = 80$), predominantly from surface waters, along the eastern Australian coast spanning Brisbane to the Tasman Sea and including near-shore waters of the East Australian Current (EAC). These water bodies comprise strong latitudinal gradients of temperature (~15-23ºC) and nutrients (0-3 µM dissolved nitrate, $NO_3^-$), moderated by eddies generated by the EAC that transiently incur nutrient-rich waters onto the continental shelf (Oke and Middleton, 2001). We examine the extent to which variance of $\Phi_{e,C}$ can be explained by corresponding changes of environmental condition. Furthermore, based on recent observations from contrasting water types in the South China Sea (Zhu et al. 2016, 2017) and eastern Australian coast (Robinson et al. 2014 Hughes et al. 2018b), we further tested how well $\Phi_{e,C}$ variance could also be explained by a broad descriptor of the capacity for phytoplankton to acquire and utilise resources. Specifically, the predominant cell size fraction (measured as size-fractionated Chl-*a*) was chosen, since it is routinely incorporated into oceanographic studies as a broad identifier of community composition. Finally, as recent field studies of $\Phi_{e,C}$ have shown a strong relationship between NPQ and $\Phi_{e,C}$ (Schuback et al. 2015, 2016, 2017; Schuback and Tortell, 2019; Hughes et al. 2018b) we also examined how well this photophysiological metric could predict $\Phi_{e,C}$ variance across a dynamic nutrient regime and whether a new method proposed for baseline correction (Boatman et al. 2019) might explain such variance.

## 2. Materials and Methods

*2.1. Study area and sample collection*

A total of 80 samples were obtained for coupled FRRf and $^{14}$C-uptake measurements from along the eastern coast of Australia between August 31$^{st}$ - September 22$^{nd}$ 2016, from the *RV Investigator* (voyage: IN2016_v04). Discrete seawater samples were collected from surface waters (5-7 m depth), and from the sub-surface chlorophyll maximum (SSCM) when discernible, via Conductivity Temperature Depth (CTD) casts (SBE32, Seabird Electronics, USA), using a 24 bottle rosette sampler. Presence of SSCMs was identified from vertical Chl-*a* fluorescence profiles from a passive fluorometer attached to the CTD frame. Additional surface samples were collected underway from the continuously-pumped surface seawater supply system (7 m intake depth, non-filtered supply). In total, *n* = 64 surface samples (21 discrete, 41 underway) and *n* = 16 discrete SSCM samples were collected (Fig. 1).

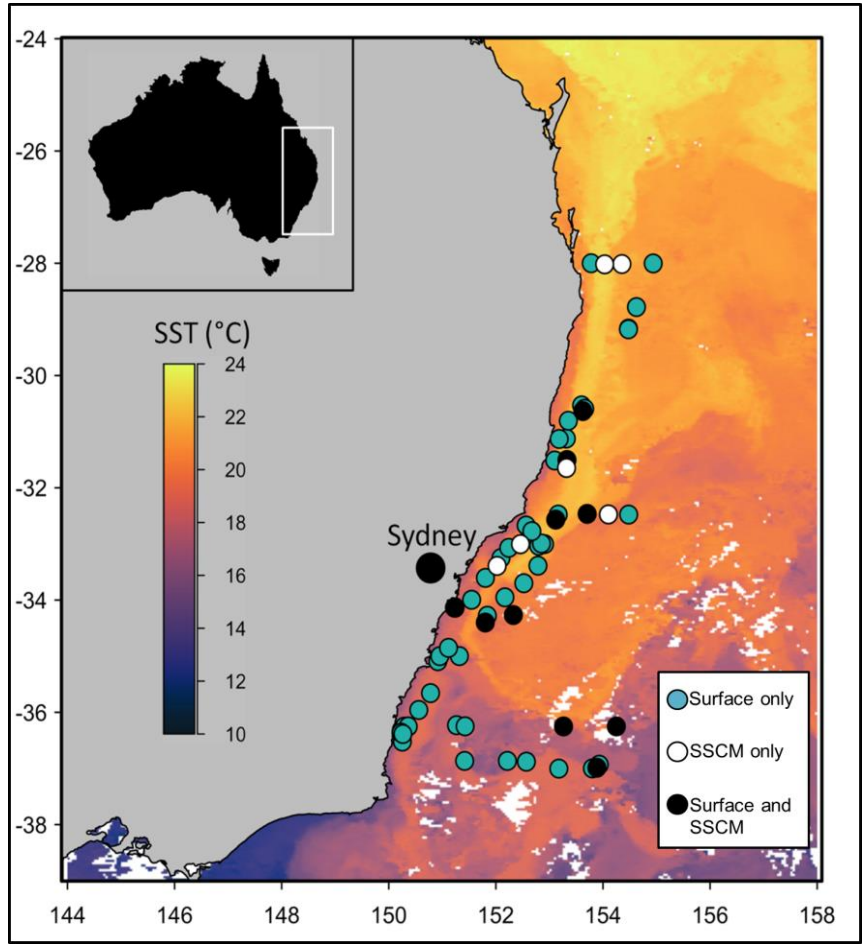

**Figure 1** Study area and sampling locations. Green markers denote locations at which only surface samples were collected, while white markers indicate locations at which only the sub-surface Chlorophyll-*a* maximum (SSCM) was sampled via Conductivity Temperature Depth casts (CTD). Black markers indicate locatiosn where both the SSCM and surface was sampled ($n = 80$ total samples). Sea surface temperature (SST) at the beginning of the IN2016_v04 research voyage (22$^{nd}$ August, 2016) is overlaid (data sourced from the Integrated Marine Observing System, IMOS data portal: http://imos.aodn.org.au).

*2.2. Physico-chemical parameters*

Continuous measurements of photosynthetically active radiation (PAR) were taken in air using two quantum sensors (Li-190, Li-COR, Lincoln, NE, USA) positioned on the starboard and port sides of the upper deck. Salinity and temperature were determined for each discrete sample from probes attached to the CTD sensor (SeaBird SBE911 dual conductivity and temperature sensor) or the pumped underway seawater supply (Seabird SBE21 SeaCAT thermosalinograph). Analysis of dissolved inorganic nutrient concentrations: ammonium ($NH_4^+$), nitrate ($NO_3^-$), phosphate ($PO_4^{3-}$) and silicate (Si) was conducted at sea immediately after sample collection (for both underway and discrete samples), following protocols outlined in Ajani et al. (2018). Analysis was performed using a SEAL AA3HR segmented flow injection analyser with instrument detection limits of: 0.02, 0.2, 0.02 and 0.2 µM for $NH_4^+$, $NO_3^-$, $PO_4^{3-}$ and Si respectively. Dissolved organic carbon (DIC) was immediately quantified onboard from the mean of triplicate measurements using an Apollo SciTech dissolved gas analyser (Model: AS-C3, Apollo SciTech, Newark, DE, USA).

*2.3. Size-fractionated Chl-a*

Total Chl-*a* content was determined by filtering 250 mL of seawater under low vacuum (<50 mg Hg) through a Sterlitech GF/F filter (0.3 µm nominal pore size). Filters were then transferred to 20 mL glass vials and pigments were immediately extracted using 3 mL 90% acetone and stored in the dark at 4°C for 24 hours. For the 2-10 µm and >10 µm fractions, a similar procedure was performed, using 2 µm Glass-fibre filters (Microanalytix, Sydney, Australia) and 10 µm polycarbonate filters (Merck Millipore, Bayswater, VIC, Australia) respectively. Chl-*a* was then determined fluorometrically using a Trilogy fluorometer (Turner Designs, California, USA, serial number: 720000354), equipped with a non-acidification Chl-*a* module (Turner Designs, USA) and calibrated against a pure Chl-*a* standard (Sigma-Aldrich Pty Ltd, Castle Hill, NSW, Australia).

*2.4. Photophysiological characterisation and electron transport rates (FRRf)*

Samples were maintained under very low light (2-3 µmol photons m$^{-2}$ s$^{-1}$) for a minimum of 20 minutes to relax non-photochemical processes prior to FRRf measurements. A FastOcean MKIII FRRf coupled to a FastAct laboratory system (Chelsea Technologies Group, London, UK) was programmed to deliver single turnover (ST) saturation of PSII from a succession of 100 flashlets (1 µs pulse with a 2 µs interval between flashes), followed by a relaxation phase of 40 flashlets (1 μs pulse with a 50 μs interval between flashes). A total of 100 sequences were performed per acquisition, with an interval of 150 ms between sequences. For all ST measurements, the blue LED (450nm) was the sole excitation source used to drive closure of PSII reaction centres and generate fluorescence induction transients. The biophysical model of Kolber et al. (1998) was fitted to all fluorescent transients using FastPro8 software (V.1.0.55; Chelsea Technologies) to determine minimum ($F_0$, $F'$) and maximum fluorescence ($F_m$, $F_m'$), functional absorption cross section of PSII ($\sigma_{PSII}$, $\sigma_{PSII}'$) and the PSII connectivity factor ($\rho$, $\rho'$) (where the prime notation denotes that samples were measured during exposure to actinic light). Contribution of background fluorescence emitted from fluorescent dissolved organic matter was measured from 0.2 μm-filtered samples and subsequently subtracted from all samples within the FastPro8 software. For this study we calculated $v\text{ETR}_{PSII}$ (mol electrons m$^{-3}$ s$^{-1}$) according to the "absorption" algorithm of Oxborough et al. (2012) (Eq. 1):

$$v\text{ETR}_{PSII} = a_{LHII} \cdot \frac{F_q'}{F_m'} \cdot E \tag{1}$$

Where $a_{LHII}$ is the absorption coefficient for PSII light harvesting (units: m$^{-1}$), $F_q'/F_m'$ is the effective PSII quantum yield under ambient light (Genty et al. 1989) and $E$ is irradiance (mol photons m$^{-2}$ s$^{-1}$). The absorption algorithm represents a modified version of the "sigma" algorithm (Kolber et al. 1998), which allows for parameterisation of $a_{LHII}$ without $\sigma_{PSII}'$ that can be difficult to resolve reliably under high ambient irradiance (see Oxborough et al. 2012) (Eq. 2):

$$a_{LHII} = \frac{F_m \cdot F_0}{F_m - F_0} \cdot K_a \tag{2}$$

Where $K_a$ is an instrument-specific constant (units: m$^{-1}$). This approach does not require an assumption of the connectivity of PSII reaction centres (RCIIs) (see Kolber et al. 1998; Kramer et al. 2004) and thus the absorption algorithm can be alternatively expressed as per

Eq. 3. Strictly, this derivation is denoted $JV_{PSII}$, since it is a flux (see Oxborough et al. 2012) but for consistency with terminology in Eqs 1, we continue to refer to this as $vETR_{PSII}$:

$$vETR_{PSII} = \frac{F_m \cdot F_0}{F_m - F_0} \cdot \frac{F_{q'}}{F_{m'}} \cdot K_a \cdot E \tag{3}$$

Measurements of $a_{LHII}$ are spectrally-weighted towards the FRRf excitation LED (450 nm for the instrument used in this study) and therefore were spectrally-adjusted using a spectral correction factor (SCF) informed by knowledge of the dominant phytoplankton class present in each sample, as assessed by on-board microscopy. Once the dominant phytoplankton group was identified, we followed the procedure of Hughes et al. 2018b and used previously-collected fluorescence excitation spectra (400–700 nm) from phytoplankton cultures pre-treated with 3-(3,4-dichlorophenyl)-1,1-dimethylurea (DCMU) to obtain spectrally-resolved values of $a_{LHII}(\lambda)$ as:

$$a_{LHII}(\lambda) = \left(a_{LHII}(450) \big/ F_{730}(450)\right) \cdot F_{730}(\lambda) \tag{4}$$

Values of $a_{LHII}(\lambda)$ were then adjusted to the spectral output of the actinic light source housed within the FRRf optical head as per Eq. 5.

$$\overline{a_{LHII}(\lambda)} = (\sum_{400}^{700} a_{LHII}(\lambda) \cdot E(\lambda))\Delta\lambda / \sum_{400}^{700} E(\lambda)\Delta\lambda \tag{5}$$

A total of four dominant phytoplankton classes were observed in this study (chlorophytes, diatoms, dinoflagellates and haptophytes), and SCFs applied to the data ranged from 0.37 – 0.44 (see Supplementary Fig. S1 for phytoplankton fluorescence excitation spectra used to derive SCFs). For 11 out of 80 samples there were insufficient (< 100) cells counted to allow reliable determination of taxonomic composition. For these samples, routine assessment of phytoplankton pigment group contributions to the FRRf signal was performed using a second custom multi-spectral FRRf (Soliense Inc, California, USA, no serial number). To screen for pigment contributions to $\sigma_{PSII}$, the Soliense Inc. FRRf was programmed to deliver "Flash, Length, Delay, Inc" of "100, 1.6, 5, 1 (excitation) and "80, 1.6, 20, 1.06 (relaxation), sequentially cycling through three excitation LED wavelengths: 445 nm, 470 nm and 505 nm, with acquisitions averaged from 20-80 sequences (depending upon biomass). We then compared PSII fluorescence (as $F_o$) at each wavelength (normalised to 445 nm) against the fluorescence excitation spectra for each of the four dominant phytoplankton groups (also normalised to 445 nm) – selecting the "best match" phytoplankton group as the one with the lowest cumulative difference to our sample (see Supplementary Fig. S2). We acknowledge that applying SCFs based on phytoplankton class assessed purely by microscopy (or by FRRf pigment contribution) provides limited taxonomic resolution by not accounting for co-dominance of phytoplankton groups or potential contribution of picophytoplankton, and may therefore introduce a degree of uncertainty into our reported $\Phi_{e,C}$ values. Overall however, we consider this an improvement over not applying a SCF altogether which have been reported to introduce as much as 100-200% uncertainty in FRRf measurements of $\Phi_{e,C}$ (Silsbe et al. 2015).

Fluorescence light curves (FLCs) were performed to determine the light intensity for saturated electron transport ($E_K$) as a means to standardise irradiance values for subsequent incubations used to derive $\Phi_{e,C}$ (see following section). For this, we used a similar protocol and instrument settings as previously described in Suggett et al. (2015), with the exception that each light step was held for only 20 s to minimise the overall duration of the FLC and allow for increased sampling frequency. This protocol was consistently applied to all samples however as the FLC duration approximated a "rapid" light protocol, it is likely that steady-state fluorescence was not reached during the period of exposure (e.g. Ralph and Gademann, 2005). Parameterisation of each generated FLC was achieved by fitting the model of Platt et al. (1981) to the photosynthesis-irradiance response: i.e. $\nu ETR_{PSII}$ versus $E$ data. Non-linear curve fits were performed using Sigmaplot v11.0, (Systat Software Inc, California, USA). Least squares non-linear regression analysis of the model fit was performed to estimate the maximum rate of electron transport, $v\text{ETR}_{\text{PSII}}^{\text{max}}$, the light utilisation efficiency, α (electrons m$^{-3}$ s$^{-1}$) and thus the light saturation parameter, $E_K$ (calculated as $v\text{ETR}_{\text{PSII}}^{\text{max}}/\alpha$) with units of µmol photons m$^{-2}$ s$^{-1}$. A white LED was used to generate the FLC (hence $E_K$ is strictly weighted to this spectra, as per Moore et al. 2006), but also to subsequently illuminate samples for the FRRf-$^{14}$C productivity comparisons.

*2.5. High-throughput FRRf-$^{14}$C incubations ($\Phi_{e,C}$)* A total of 80 small-volume (3 mL), incubations were performed, where $v\text{ETR}_{\text{PSII}}$ and $^{14}$C-uptake were measured simultaneously upon the same sample (i.e. a "dual incubation") at an irradiance approximating the light-saturation parameter (described below) and $\text{ETR}_{\text{PSII}}$ was measured every 5 s during this period. To quantify $^{14}$C-uptake, we adopted the small-volume method of Lewis and Smith (1983) with several modifications. Aliquots of 3 mL seawater samples were transferred to a borosilicate test-tube and spiked to a final concentration of 0.4 µCi mL$^{-1}$ NaH$^{14}$CO$_3$ (Perkin-Elmer, Melbourne, Australia). The test-tube containing the radiolabelled sample was then incubated for 2 hr within the FRRf optical head, using the FRRf's in-built white LED array to provide actinic light. Such an approach avoids numerous errors associated with artefacts introduced by use of separate incubations (see Suggett et al. 2009a; Lawrenz et al. 2013) yet does not allow for replication as the instrument can only hold a single test tube at a time; previous assessments of laboratory cultures however demonstrated small variability between biological replicates when measured at saturating irradiance (Hughes, 2018). Upon completion of the incubation, samples were removed and immediately acidified with 150 µL of 6 M HCl to convert remaining unfixed inorganic $^{14}$C to $^{14}$CO$_2$. Samples were then de-gassed for 24 hr in a fume hood before fixation with 10 mL scintillation fluid (Ultima Gold LLT, Perkin Elmer). Fixed samples were then boxed and stored in a cool location for analysis upon return to the University of Technology Sydney (UTS). At UTS, $^{14}$C samples were shaken vigorously for several minutes and left to stand overnight before three rounds of liquid scintillation counting (count time: 5 min) using a Tri-Carb 2810 TR, Perkin-Elmer), with automatic quench correction. Rates of $^{14}$C-fixation (mol C m$^{-3}$ hr$^{-1}$) were then calculated from the concentration of DIC and the quantity of $^{14}$C isotope incorporated during the incubation as per Knap et al. (1996). $v$ETR$_{PSII}$ was scaled to hourly-integrated rates as per Suggett et al. (2009a) to allow determination of $\Phi_{e,C}$ (mol e$^-$ [mol C]$^{-1}$) as:

$$\Phi_{e,C} = \frac{v\text{ETR}_{PSII}}{C-fixation} \qquad (6)$$

Values of $\Phi_{e,C}$ have been shown to increase under saturating light intensities (i.e. the $v$ETR$_{PSII}$$^{max}$ region of a PE curve) compared to light-limiting irradiances (the α region of a PE curve) (Brading et al. 2013). Therefore, to standardise incubation conditions between samples, we opted to provide the irradiance level corresponding to the measured light-saturation parameter ($E_K$) for each sample rather than *in-situ* light levels (not measured in this study), since $E_K$ provides a convenient indicator of photoacclimation status (Sakshaug et al. 1997). Due to constraints of instrumentation in the onboard radiation laboratory, $E_K$ values were initially obtained from raw RLC data using FastPro8 software. Upon initial determination of $E_K$, the FRRf white LED array was then programmed to deliver the closest corresponding irradiance level available from a range of pre-defined irradiances allowed by the instrument software. All FLC data were however later exported for quality-controlling and subsequent fitting of the Platt et al. (1981) model using Sigmaplot to re-calculate all $E_K$ values as described above. Chosen incubation irradiances were subsequently expressed relative to the re-calculated $E_K$ values at a later time as $E/E_K$ (dimensionless). Even so, the incubation irradiances chosen at the time of sampling were very close to the re-calculated $E_K$ value (mean $E/E_K$ = 1.15), with $E/E_K$ values ranging from 0.7 – 1.7 across all samples (data



not shown). Thus, the incubation irradiances in this study represent a reasonably-constrained continuum of light-limited ($E/E_K$ <1) to light-saturated ($E/E_K$ >1) conditions for photosynthesis.

*2.6. $NPQ_{NSV}$ and PSU size*

Non-photochemical quenching (NPQ) of fluorescence was calculated as the normalised Stern-Volmer coefficient (denoted here as $NPQ_{NSV}$) according to McKew et al. (2013) as:

$$NPQ_{NSV} = \frac{1}{F_v'/F_m'} - 1 \qquad (7)$$

$NPQ_{NSV}$ was measured every 5 s during the simultaneous $^{14}$C-FRRf incubations and values reported correspond to the average of the last three measurements taken after a period of 30 min. We chose this length of time as previous studies (e.g. Schuback et al. 2016) have calculated $NPQ_{NSV}$ from measurements performed during FLCs where samples have been exposed to light for between 3-30 min. Indeed, we observed no differences in $NPQ_{NSV}$ calculated from measurements after 3 and 30 min, yet saw a significant increase by the end of the incubation period (120 min) (Supplementary Fig. S3). The size of the photosynthetic unit (PSU, with units of mol Chl-*a* [mol RCII]$^{-1}$) was derived from the fluorometric estimate of RCII according to Oxborough et al. (2012) (see also Murphy et al. 2017) and the total Chl-*a* concentration for each sample determined fluorometrically using a Triology fluorometer as outlined above.

*2.7. Statistical Analysis*

All multivariate analysis was performed using the software package, PRIMER v6 (PRIMER-E, Plymouth, UK). Multidimensional scaling (MDS) plots were constructed to visualise patterns in physico-chemical variables. Hierarchical cluster analysis (HCA) with a SIMPROF test ($p = 0.05$) was performed on a Euclidean resemblance matrix of physico-chemical



variables (PAR, temperature, salinity, $NH_4^+$, $NO_3^-$, $PO_4^{3-}$ and Si) to identify groupings of similar hydrography. Student's t-tests were used to test for statistical differences between data clusters identified by HCA for variables where assumptions of normality and equal variance were met, (tested by the Kolmogorov-Smirnov and Levene's test respectively) using Sigmaplot v11.0 (Systat Software Inc, California, USA). Where one or both assumptions were not met, differences between clusters were assessed using the Mann-Whitney Rank Sum test (Sigmaplot v11.0, Systat Software Inc.). Distance-based linear modelling (DistLM) was performed to examine how much variability in $\Phi_{e,C}$ could be explained by core environmental variables (PAR, temperature, salinity and nutrients), $E/E_K$, total Chl-*a*, size-fractionated Chl-*a* and $NPQ_{NSV}$ within clusters identified by HCA and for all data pooled. To understand the respective predictive power of the measured variables, we initially performed DistLM analysis selecting only core environmental variables as available predictor variables. We then sequentially included $E/E_K$, total Chl-*a*, size-fractionated Chl-*a* and $NPQ_{NSV}$ as additional available predictor variables, repeating DistLM analyses after addition of each variable. To obtain the most parsimonious model at each step, we used the best model selection routine based on 9999 permutations with Akaike information criterion (AICc - corrected for small sample number), which incorporates a penalty factor for increasing the number of predictor variables (Anderson et al. 2008).

Prior to DistLM, the distribution of each physico-chemical variable was assessed using draftsman's plots and co-correlations were identified from Pearson's correlation matrices. Variables with skewed distribution were square-root transformed, and if pairs of variables had a Pearson's correlation co-efficient of >0.8, one of the pair was excluded from subsequent analyses. Distance-based redundancy analysis (dbRDA) plots were generated to enable two-dimensional visualisation of the best DistLM models. Significant differences between $NPQ_{NSV}$ measured after variable incubation lengths (3, 30 and 120 min) and the



mean $NPQ_{NSV}$ over the entire incubation were evaluated using the non-parametric Kruskal-Wallis test followed by Dunn's post hoc test. An identical procedure was also used to assess statistical differences between $\Phi_{e,C}$ when binned into groups based on dominant Chl-*a* size fraction.

## 3. Results

*3.1. Physico-chemical characterisation*

Sampled water masses were characterised by a distinct gradient of both nutrients and temperature (Fig. 2a-f). $NO_3^-$, $PO_4^{3-}$ and Si exhibited higher concentrations in coastal samples and the southern Tasman Sea, compared to the EAC and northernmost coastal/oceanic samples (Fig. 2b-d). $NH_4^+$ concentrations were generally highest in coastal samples (up to 0.6 µM), but relatively low in the Tasman Sea and EAC, where values occasionally fell below the limit of detection (0.02 µM; Fig. 2a). Temperature exhibited a distinct latitudinal pattern, ranging from ~15ºC in the southern Tasman Sea to ~23.5ºC for northernmost coastal samples, with a distinct thermal gradient also measured within the EAC, where surface temperatures cooled with southward travel (Fig. 2f). Salinity remained largely consistent throughout (~35-35.8 ppt), yet the highest values were consistently measured within the EAC water mass (Fig. 2e).

*3.2. $\Phi_{e,C}$, biomass and photophysiology*

Measured $\Phi_{e,C}$ ranged from ~4.7 to 65 mol e$^-$ (mol C)$^{-1}$, with a mean of ~16 mol e$^-$ (mol C)$^{-1}$ during this study (Fig. 3a). Thus, all measurements of $\Phi_{e,C}$ were higher than the theoretical minimum value of 4 mol e$^-$ (mol C)$^{-1}$, and the upper values agreed well with a previous meta-analysis of global FRRf-derived $\Phi_{e,C}$ data (Lawrenz et al. 2013) and recent field campaigns (e.g. Ko et al. 2019). $\Phi_{e,C}$ was generally lowest in coastal waters, where virtually all values



fell below the observed mean for this study (i.e. <16 mol e$^-$ [mol C]$^{-1}$). The EAC and the Tasman Sea were both characterised by large variability in $\Phi_{e,C}$ (values ranging from ~10-65 mol e$^-$ [mol C]$^{-1}$), and unlike physico-chemical variables measured, no latitudinal pattern in $\Phi_{e,C}$ was evident (Fig. 3a). Carbon assimilation rate per unit Chl-*a* (carbon assimilation number, measured at saturating irradiance) ranged from 0.5 – 5.5 (mean: 2.2) mg C (mg Chl-*a*)$^{-1}$ hr$^{-1}$ and was generally higher (>3 mg C [mg Chl-*a*]$^{-1}$ hr$^{-1}$) in coastal waters and the southern Tasman Sea, whilst lower values (<2.5 mg C [mg Chl-*a*]$^{-1}$ hr$^{-1}$) were consistently measured within the EAC (Fig. 3b).



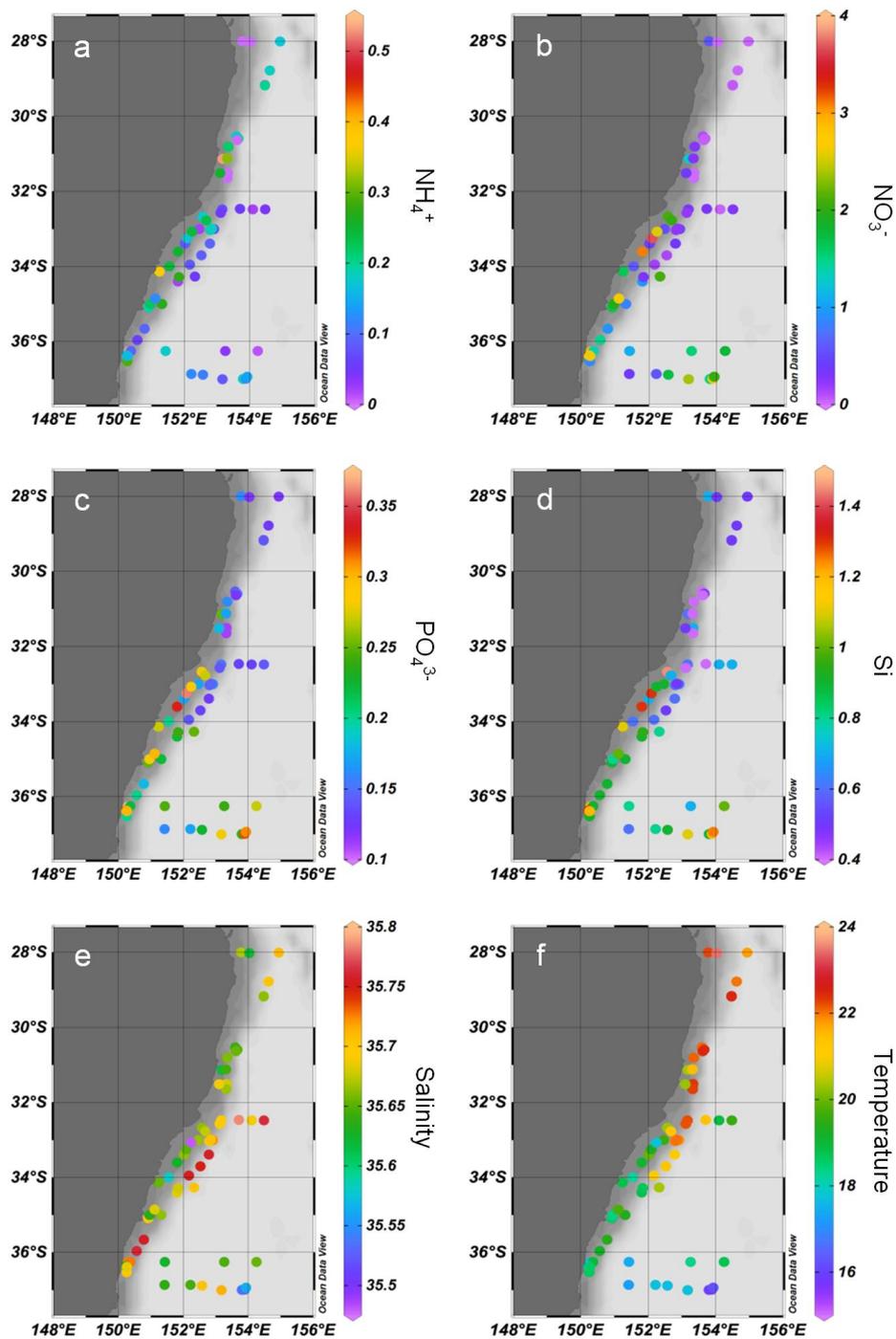

**Figure 2** Physicochemical characteristics of surface water sampled in coastal, Eastern Australian Current (EAC) and Tasman Sea water masses measured from the RV *Investigator* (August – September 2016, IN2016_v03); **(a)** ammonium ($NH_4^+$), **(b)** nitrate ($NO_3^-$), **(c)** phosphate ($PO_4^{3-}$), **(d)** silicate (Si) [all nutrient concentrations are reported as μM], **(e)** salinity (ppt) and **(f)** sea surface temperature (°C).



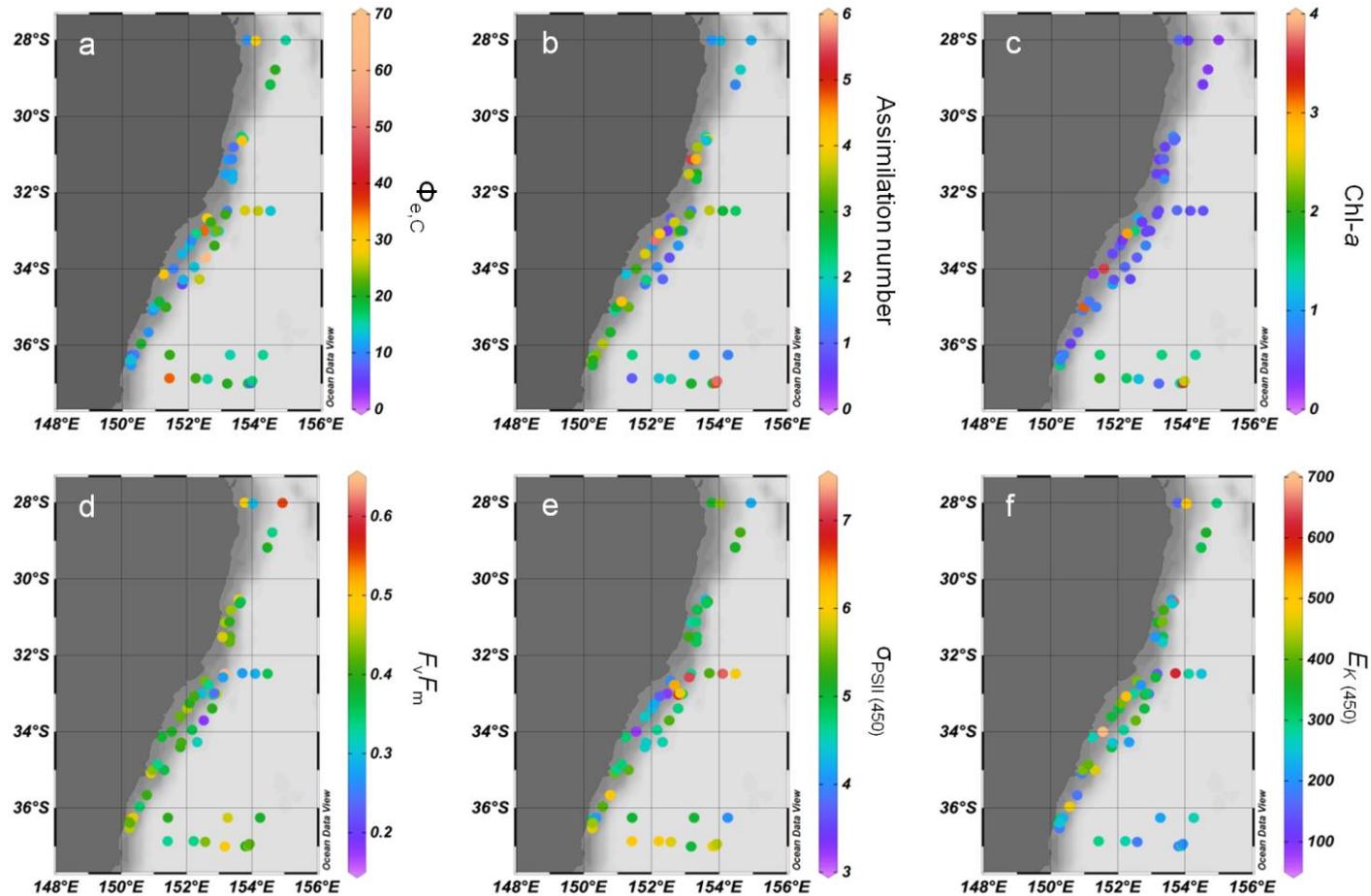

**Figure 3** Photophysiological and productivity characteristics of surface water sampled in coastal, Eastern Australian Current (EAC) and Tasman Sea water masses measured from the RV *Investigator* (August – September 2016, IN2016_v04); (a) the electron requirement for carbon fixation ($\Phi_{e,C}$, mol e$^-$ [mol C]$^{-1}$), (b) carbon assimilation number (mg C [mg Chl-*a*] hr$^{-1}$), (c) Chlorophyll-*a* (Chl-*a*) biomass (mg/m$^3$), (d) maximum photochemical efficiency ($F_v/F_m$, dimensionless), (e) functional absorption cross-section of PSII ($\sigma_{PSII}$, nm$^2$ PSII$^{-1}$) and (f) light-saturation parameter ($E_{K[450]}$, µmol photons m$^{-2}$ s$^{-1}$).



Total Chl-*a* biomass ranged from 0.1 to 3.5 mg m$^{-3}$, averaging ~0.8 mg m$^{-3}$ during the study, with highest Chl-*a* concentrations occurring within coastal waters and the southernmost Tasman Sea (Fig. 3c). The 2-10 µm Chl-*a* size fraction was the most dominant size fraction in this study, representing the largest contributor to total Chl-*a* in nearly half (42%) of all samples (Fig. 4a). Conversely, the <2 µm size fraction was least dominant, accounting for the majority of total Chl-*a* in only 15 out of 80 samples (19%), with the >10 µm size-fraction intermediate (~30% of samples) (Fig. 4a). Notably, $\Phi_{e,C}$ did not appear to exhibit an obvious pattern relating to the dominant Chl-*a* size fraction contributing to total Chl-*a* (Fig. 4a)

Values for the maximum photochemical efficiency ($F_v/F_m$) ranged from 0.24 to 0.57 and followed a similar pattern to Chl-*a*, with larger values measured in coastal waters and the southern Tasman Sea (Fig. 3d). Conversely, the functional absorption cross-section of PSII ($\sigma_{PSII}$), was generally lowest in coastal waters (albeit with a degree of variability), although the largest values recorded (>6 nm$^2$ PSII$^{-1}$) corresponded to the southern Tasman Sea water mass (Fig. 3e).

FLC-retrieved values of the light saturation parameter ($E_K$) spanned a wide range of values (~85-700 µmol photons m$^{-2}$ s$^{-1}$) with a mean value of ~270 µmol photons m$^{-2}$ s$^{-1}$. $E_K$ was consistently low (< 400 µmol photons m$^{-2}$ s$^{-1}$) in both the southern Tasman Sea water mass and the EAC (with the exception of a single sample), yet exhibited far greater variability in coastal waters (Fig. 3f).



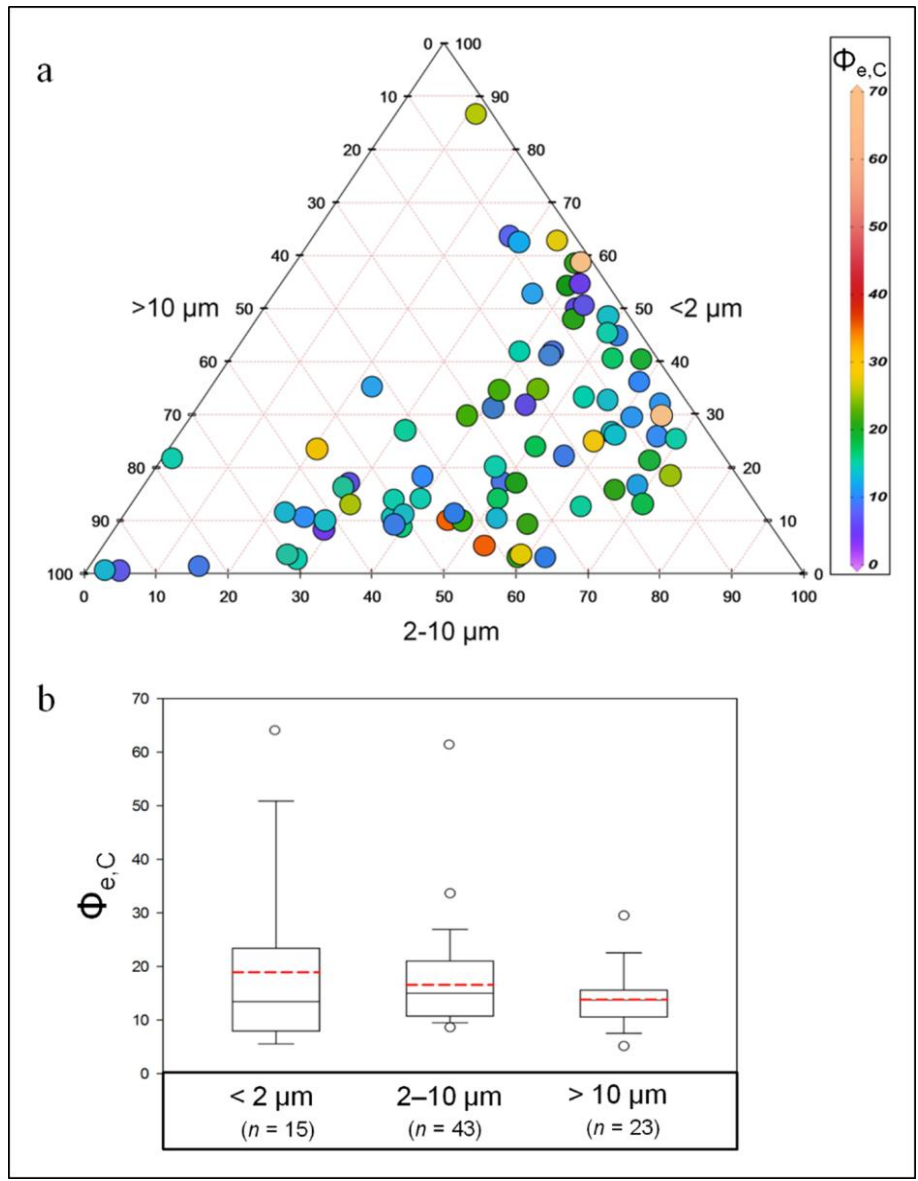

**Figure 4** a) Simplex plot of the electron requirement for carbon fixation, $\Phi_{e,C}$ (mol e$^-$ [mol C]$^{-1}$) showing % contribution of Chl-*a* size fractions (<2 µm, 2-10 µm and >10 µm) for individual samples (*n* = 80); b) Box-plot of $\Phi_{e,C}$ binned according to the dominant Chl-*a* size fraction (*n* = 81, note that one sample appears in both the <2 µm and 2-10 µm fractions due to equal dominance of both size fractions). For each cluster the median and mean values are indicated by the bold and red dashed lines respectively, the large box represent the 5$^{th}$ and 95$^{th}$ percentiles and the open circles denote outliers.



*3.3. Predicting $\Phi_{e,C}$ from physico-chemical versus taxonomic variables*

Previous studies modelling $\Phi_{e,C}$ variability have demonstrated improved predictive capacity by grouping samples according to similar hydrography (e.g. Lawrenz et al. 2013; Zhu et al. 2016, 2017), therefore we adopted a similar approach here. MDS identified a distinct separation of data based on nutrients and temperature (Fig. 5a), with HCA identifying two primary clusters (Supplementary Fig. S4), predominately separated latitudinally (Fig. 5b): **Cluster A)** generally characterised by higher temperatures and lower dissolved nutrient levels, and **Cluster B)** characterised by lower temperatures and higher dissolved nutrient levels (Table 1). Overall, 12 of the 16 SSCM samples were assigned to cluster B (data not shown). Binning these various data according to the two clusters revealed broad differences in the physiology inherent to the two prevailing water types sampled:

Mean $\Phi_{e,C}$ was lower and carbon assimilation number was higher in cluster A (15.1 mol e$^-$ [mol C]$^{-1}$) and 2.6 mg C [mg Chl-*a*]$^{-1}$ hr$^{-1}$ respectively) compared to cluster B (18.96 mol e$^-$ [mol C]$^{-1}$) and 2.11 mg C [mg Chl-*a*]$^{-1}$ respectively) (Table 1). Cluster B was further characterised by significantly higher total Chl-*a* biomass, comprised of a greater proportion of large cells as inferred from >10μm Chl-a size fraction (35% vs. 20%) (Mann-Whitney Rank sum test, Table 1). Mean PSU size was similar between clusters (Table 1, Supplementary Fig. S5), with a value of ~600 mol Chl-*a* (mol RCII)$^{-1}$ - 20% larger than the commonly used reference value for eukaryotic-dominated phytoplankton assemblages (500 mol Chl-*a* [mol RCII]$^{-1}$) (Kolber and Falkowski, 1993; but see also Raateoja et al. 2004).



|  | Cluster **A** ($n = 33$) | Cluster **B** ($n = 47$) | test statistic |
|---|---|---|---|
| **Physicochemical** | | | |
| Temperature (°C) | 21.45 (0.22) | 18.60 (0.20) | $p = < 0.001^{(MW)**}$ |
| Salinity (ppt) | 35.69 (0.01) | 35.65 (0.01) | $p = 0.002^{(S)*}$ |
| $NH_4^+$ (μM) | 0.09 (0.02) | 0.17 (0.02) | $p = < 0.001^{(MW)**}$ |
| $NO_3^-$ (μM) | 0.24 (0.03) | 1.82 (0.12) | $p = < 0.001^{(MW)**}$ |
| $PO_4^{3-}$ (μM) | 0.14 (0.01) | 0.26 (0.01) | $p = < 0.001^{(MW)**}$ |
| Si (μM) | 0.56 (0.02) | 1.01 (0.03) | $p = < 0.001^{(MW)**}$ |
| **Biological** | | | |
| Chl-*a* < 2 μm (%) | 35.23 (2.81) | 20.01 (2.52) | $p = < 0.001^{(MW)**}$ |
| Chl-*a* 2-10 μm (%) | 42.85 (2.14) | 44.55 (2.70) | $p = 0.346^{(MW)}$ |
| Chl-*a* > 10 μm (%) | 21.91 (3.42) | 35.44 (3.71) | $p = < 0.001^{(MW)**}$ |
| Total Chl-*a* (mg m$^3$) | 0.63 (0.06) | 1.05 (0.13) | $p = 0.02^{(MW)*}$ |
| $F_v/F_m$ (unitless) | 0.39 (0.02) | 0.42 (0.01) | $p = 0.124^{(MW)}$ |
| $\sigma_{PSII\,(450)}$ (nm$_2$ PSII$^{-1}$) | 5.40 (0.16) | 5.12 (0.13) | $p = 0.168^{(S)}$ |
| $E_{K\,(450)}$ (μmol photons m$^{-2}$ s$^{-1}$) | 295.62 (21.77) | 269.42 (20.36) | $p = 0.257^{(MW)}$ |
| $E/E_K$ (unitless) | 1.16 | 1.15 | $p = 0.88^{(S)}$ |
| $NPQ_{NSV}$ (unitless) | 1.96 (0.78) | 1.64 (0.58) | $p = 0.135^{(MW)}$ |
| $\Phi_{e,C}$ (mol e$^-$ [mol C]$^{-1}$) | 18.96 (1.97) | 15.1 (1.39) | $p = 0.044^{(MW)*}$ |
| Carbon assimilation (mg C [mg Chl-*a*] hr$^{-1}$) | 2.11 (0.18) | 2.60 (0.18) | $p = 0.069^{(S)}$ |
| PSU size (mol Chl-*a* [mol RCII]$^{-1}$) | 617.33 (26.24) | 600.83 (28.55) | $p = 0.695^{(S)}$ |

**Table 1** Mean (± SE, standard error) of physicochemical variables and biological parameters within Cluster A and B (see Supplementary Fig. S4 for cluster information). * and ** denotes significance levels of 0.05 and <0.01 respectively assessed by either Student's t-test (denoted by $^{(S)}$) or Mann-Whitney Rank sum test denoted by $^{(MW)}$.

Changes in core environmental variables (temperature, salinity, $NH_4^+$, $NO_3^-$, $PO_4^{3-}$, Si and PAR) explained only 19% of $\Phi_{e,C}$ variation across the entire dataset (Table 2), indicating that prevailing environmental conditions were not strong predictors of $\Phi_{e,C}$ in this study area. When separated by cluster, the ability to explain variance in $\Phi_{e,C}$ improved to 22% for cluster A, but decreased for cluster B to 10% (Table 2). Including $E/E_K$ as an available predictor



variable resulted in a slight improvement of $\Phi_{e,C}$ variance explained by the best model for all data combined (to 23 %), yet was not selected in the models that best explained variability for cluster A or B individually (Table 2). Knowledge of total Chl-*a* resulted in only a marginal improvement to explained variance of $\Phi_{e,C}$ within cluster A (from 22 to 24%) yet did not feature in predictor variables that best explained variance within the combined dataset. Similarly, inclusion of size-fractionated Chl-*a* as an available predictor variable resulted in no improvement of model performance for all data combined or cluster A, yet did improve $\Phi_{e,C}$ variance explained by core environmental variables from 10 to 19% in cluster B. The lack of an obvious pattern between $\Phi_{e,C}$ and the dominant Chl-a size fraction (Fig. 4a) likely explains the lack of model improvement when available as a predictor variable. However, intriguingly when $\Phi_{e,C}$ values were binned into three sample groups based on their dominant Chl-*a* size-fraction (i.e. the largest individual contributor to overall Chl-*a* pigment), mean $\Phi_{e,C}$ appeared to decrease with increasing size fraction (Fig. 4b), although no statistical differences could be determined ($p > 0.05$. Kruskal-Wallis test). $\Phi_{e,C}$ was thus lowest for assemblages dominated by cells/pigment in the >10 µm fraction (13.8 ± 1.2 mol e$^-$ [mol C]$^{-1}$, $n = 23$), increasing to 18.6 mol e$^-$ [mol C]$^{-1}$, $n = 15$ for the < 2 µm fraction, with the 2-10 µm fraction intermediate (16.1 mol e$^-$ [mol C]$^{-1}$) (Fig. 4b).



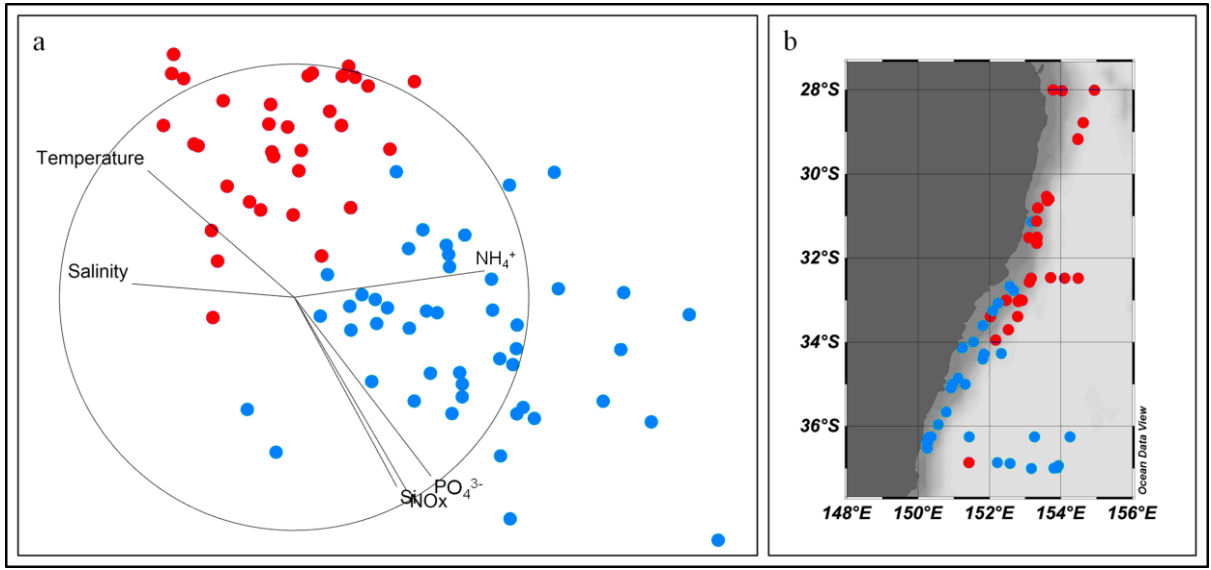

**Figure 5 a)** Multi-dimensional scaling (MDS) plot of physico-chemical (environmental) variables for samples collected from the RV *Investigator* (August – September 2016, IN2016_v04). Red and blue circles represent data clusters (A and B respectively), assigned according to hierarchical cluster analysis (CLUSTER, with SIMPROF test [p = 0.05]) performed upon a Euclidean resemblance matrix, generated from square-root transformed data; and **b)** spatial distribution of data clusters.

Finally, we included $NPQ_{NSV}$ as an available predictor variable to evaluate its potential to explain variance of $\Phi_{e,C}$ within our dataset, as has been recently demonstrated in other FRRf-based studies (e.g. Schuback et al. 2015; Hughes et al. 2018b). Linear regression showed that $NPQ_{NSV}$ in fact exhibited the strongest relationship with $\Phi_{e,C}$ of all predictor variables in this study ($r^2 = 0.51$, $p < 0.01$, Fig. 6a) with similar slopes between clusters (7.42 and 6.41 for cluster A and B respectively). Including $NPQ_{NSV}$ as an available predictor variable thus resulted in a substantial improved ability to explain $\Phi_{e,C}$ variance by the best model for all data combined (from 23% to 54%). Within Cluster A, inclusion of $NPQ_{NSV}$ within the best model yielded greatly increased ability to explain $\Phi_{e,C}$ variance, from 24% to 67%. Interestingly, although the improvement was not as large for Cluster B, the best model consisted of $NPQ_{NSV}$ as the single predictor variable, which alone explained 41% of $\Phi_{e,C}$ variance (Table 2).



| Available variables | AICc | $R^2$ | RSS | Selected variables (best) |
|---|---|---|---|---|
| **All data** ($n = 80$) | | | | |
| Core env. | 277.0 | 0.19 | 2293.6 | Temperature, $PO_4^{3-}$, PAR |
| + $E/E_K$ | 275.6 | 0.23 | 2191.1 | Temperature, $PO_4^{3-}$, PAR, $E/E_K$ |
| + Chl-*a* | 275.6 | 0.23 | 2191.1 | Temperature, $PO_4^{3-}$, PAR, $E/E_K$ |
| + S/F Chl-*a* | 275.6 | 0.23 | 2191.1 | Temperature, $PO_4^{3-}$, PAR, $E/E_K$ |
| + $NPQ_{NSV}$ | 234.7 | **0.54** | 1313.0 | Temperature, $PO_4^{3-}$, $NPQ_{NSV}$, $E/E_K$ |
| **Cluster A** | | | | |
| Core env. | 159.4 | 0.22 | 3352.9 | Temperature, $PO_4^{3-}$ |
| + $E/E_K$ | 159.4 | 0.22 | 3352.9 | Temperature, $PO_4^{3-}$ |
| + Chl-*a* | 158.4 | 0.24 | 3260.3 | $PO_4^{3-}$, Chl-*a* |
| + S/F Chl-*a* | 158.4 | 0.24 | 3260.3 | $PO_4^{3-}$, Chl-*a* |
| + $NPQ_{NSV}$ | 136.7 | **0.67** | 1433.7 | $PO_4^{3-}$, PAR, $NPQ_{NSV}$, Chl-*a* |
| **Cluster B** | | | | |
| Core env. | 228.1 | 0.10 | 5495.7 | $PO_4^{3-}$ |
| + $E/E_K$ | 228.1 | 0.10 | 5495.7 | $PO_4^{3-}$ |
| + Chl-*a* | 228.1 | 0.10 | 5495.7 | $PO_4^{3-}$ |
| + S/F Chl-*a* | 227.8 | 0.19 | 4950.7 | $PO_4^{3-}$, Chl-*a* <2μM, Chl-*a* >10μM |
| + $NPQ_{NSV}$ | 208.66 | **0.41** | 3636.3 | $NPQ_{NSV}$ |

**Table 2.** Variance in $\Phi_{e,C}$ explained by DistLM analysis using only core environmental variables (Core env.), conisisting of surface irradiance (PAR), temperature, salinity and nutrients ($NH_4^+$, $NO^{3-}$, $PO_4^{3-}$ and Si) as available predictor variables, then sequentially adding additional available predictor variables: irradiance relative to light saturation parameter ($E/E_K$), total chlorophyll-*a* content (Chl-*a*), size fractionated Chl-*a* % (<2μM, 2-10 μM and >10μM) and normalised Stern-Volmer non-photochemical quenching ($NPQ_{NSV}$). Shown are the selected variables from the best model generated after adding each new available predictor variable, together with the Akaike information criterion statistic (AICc – corrected for small sample number) and residual sum of squares (RSS). Bold text denotes variance explained by the best model generated from all available predictor variables.

*3.4. Evaluating potential influence of baseline fluorescence on $\Phi_{e,C}$ and $NPQ_{NSV}$*

Ours is the latest in a series of recent studies that have collectively demonstrated a strong relationship between $\Phi_{e,C}$ and $NPQ_{NSV}$. As the present dataset spans a dynamic environmental gradient of nutrients, and consists of a wide range of measured $F_v/F_m$ values (from 0.24 to 0.57), it presents an opportunity to examine the potential significance of baseline fluorescence for FRRf-based measures of $\Phi_{e,C}$ and its apparent strong relationship with



$NPQ_{NSV}$. Boatman et al. (2019) used coupled $O_2$-flash yield and FRRf measurements to assess potential errors in $vETR_{PSII}$ determination caused by baseline fluorescence of cellular origin (denoted as $F_b$), proposing a fluorescence-based correction procedure to compensate for this effect. Therefore, following recommendations by Boatman et al. (2019) we estimated the contribution of $F_b$ in the dark-adapted state as:

$$F_b = F_m - \frac{F_v}{0.5} \tag{8}$$

And for each sample measurement in the light-adapted state as:

$$F_b' = F_b \cdot \frac{F_m'}{F_m} \tag{9}$$

This estimation of $F_b$ contribution makes two key assumptions, specifically that: i) a value of 0.5 approximates the assumed consensus photochemical efficiency for each sample (see Boatman et al. 2019) and ii) $F_b$ is emitted from the same thylakoid membranes as $F_v$, and therefore is quenched to the same extent as $F_m$ during PSII downregulation (Oxborough et al. 2012). Calculated $F_b$ was then subtracted from dark-adapted values of $F_o$ and $F_m$ for each sample, before $a_{LHII}$ was re-calculated as per Eq. 2. Light-adapted values of $F'$ and $F_m'$ were similarly corrected by subtracting $F_b'$ to allow for recalculation of $F_q'/F_m'$, and subsequently of $vETR_{PSII}$ as per Eq. 1. For samples where the pre-corrected $F_v/F_m$ was >=0.5, no $F_b$ correction procedure was applied (as per Boatman et al. 2019).

| Available variables | AICc | $R^2$ | RSS | Selected variables (best) |
|---|---|---|---|---|
| **All data** ($n = 80$) | | | | |
| Core env. | 376.6 | 0.11 | 8185.3 | Temperature, $PO_4^{3-}$ |
| + $E/E_K$ | 373.4 | 0.17 | 7648.7 | Temperature, $PO_4^{3-}$, $E/E_K$ |
| + Chl-*a* | 373.4 | 0.17 | 7648.7 | Temperature, $PO_4^{3-}$, $E/E_K$ |
| + S/F Chl-*a* | 373.2 | 0.19 | 7417.0 | $PO_4^{3-}$, $E/E_K$, Chl-*a* <2µM, Chl-*a* >10µM |
| + $NPQ_{NSV}$ | 341.6 | **0.46** | 4999.3 | Temperature, $PO_4^{3-}$, $NPQ_{NSV}$, $E/E_K$ |
| **Cluster A** | | | | |
| Core env. | 161.6 | 0.21 | 3318.3 | Temperature, salinity, $PO_4^{3-}$ |
| + $E/E_K$ | 160.2 | 0.29 | 2925.1 | Temperature, salinity, $PO_4^{3-}$, $E/E_K$ |
| + Chl-*a* | 156.4 | 0.32 | 2839.1 | $NO_3^-$, $E/E_K$, Chl-*a* |
| + S/F Chl-*a* | 156.4 | 0.32 | 2839.1 | $NO_3^-$, $E/E_K$, Chl-*a* |
| + $NPQ_{NSV}$ | 139.3 | **0.69** | 1283.1 | $PO_4^{3-}$, $NH_4^+$, PAR, $NPQ_{NSV}$, $E/E_K$, Chl-*a* |
| **Cluster B** | | | | |
| Core env. | 216.5 | 0.09 | 4302.1 | $PO_4^{3-}$ |
| + $E/E_K$ | 216.5 | 0.09 | 4302.1 | $PO_4^{3-}$ |
| + Chl-*a* | 216.5 | 0.09 | 4302.1 | $PO_4^{3-}$ |
| + S/F Chl-*a* | 216.5 | 0.14 | 4089.3 | $PO_4^{3-}$, Chl-*a* >10µM |
| + $NPQ_{NSV}$ | 203.9 | **0.34** | 3127 | $NO_3^-$, $NPQ_{NSV}$ |

**Table 3** Variance in $\Phi_{e,C}$ explained by DistLM analysis *after correction for contribution of baseline fluorescence* ($F_b$) using only core environmental variables (Core env.), conisisting of surface irradiance (PAR), temperature, salinity and nutrients ($NH4^+$, $NO^{3-}$, $PO_4^{3-}$ and Si) as available predictor variables, then sequentially adding additional available predictor variables: irradiance relative to light saturation parameter ($E/E_K$), total chlorophyll-*a* content (Chl-*a*), size fractionated Chl-*a* % (<2µM, 2-10 µM and >10µM) and normalised Stern-Volmer non-photochemical quenching ($NPQ_{NSV}$). Shown are the selected variables from the best model generated after adding each new available predictor variable, together with the Akaike information criterion statistic (AICc – corrected for small sample number) and residual sum of squares (RSS). Bold text denotes variance explained by the best model generated from all available predictor variables.

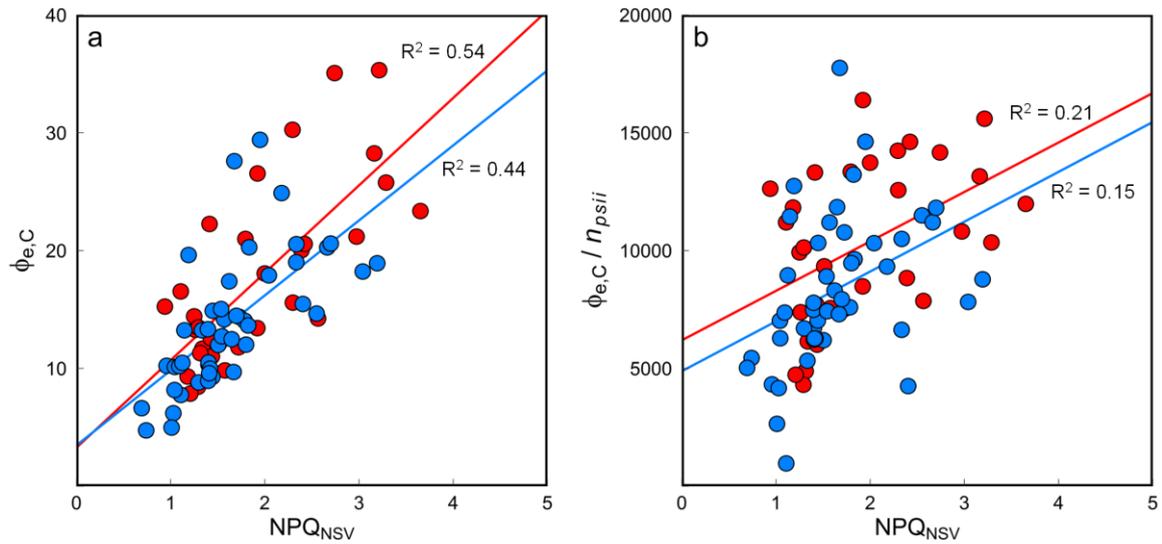

**Figure 6** Relationship between **a)** the electron requirement for carbon fixation, $\Phi_{e,C}$ (mol e$^-$ [mol C]$^{-1}$) and the expression of non-photochemical quenching (NPQ$_{NSV}$, calculated as per McKew et al. 2013) with a generated regression equation for all data combined of y = 7.186x + 2.7739 ($R^2$ = 0.51), excluding outliers >40 mol e$^-$ (mol C)$^{-1}$ (*n* = 2) and **b)** $\Phi_{e,C}$/nPSII against NPQ$_{NSV}$ (i.e. without estimation of Photosynthetic Unit [PSU] size as per Schuback et al. 2015). Data shown corresponds to clusters A and B (red circles = cluster A, blue circles = cluster B;) based on physicochemical variables (Table 1). NPQ$_{NSV}$ reflects an integrated value over a period of two hours, at an irradiance approximating the light-saturation parameter ($E_K$) for each sample.

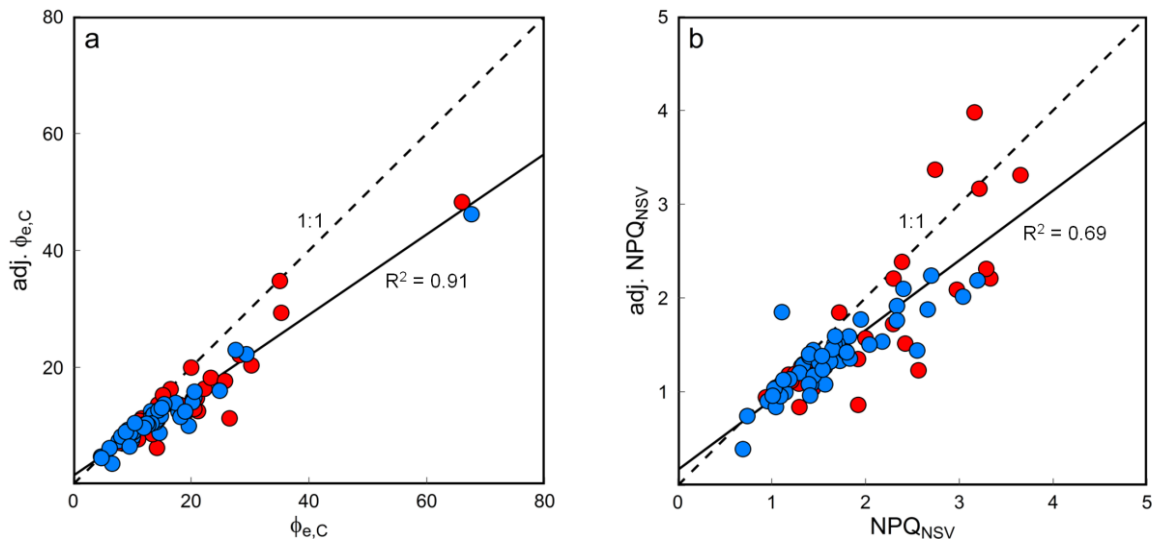

**Figure 7** Comparison of data pre- and post- correction for baseline fluorescence ($F_b$) **a)** the electron requirement for carbon fixation, $\Phi_{e,C}$ (mol e$^-$ [mol C]$^{-1}$) and **b)** the expression of non-photochemical quenching (NPQ$_{NSV}$, calculated as per McKew et al. 2013). Data shown corresponds to clusters A and B based on hydrography (red circles = cluster A, blue circles = cluster B, see Fig. 3, Table 1). NPQ$_{NSV}$ reflects an integrated value over a period of 2 hours, at an irradiance approximating the light-saturation parameter ($E_K$) for each sample.

Correcting for baseline fluorescence influence yielded a narrower range of $\Phi_{e,C}$ values (3.5 to 47 mol e$^-$ [mol C]$^{-1}$) compared to uncorrected (4.7 to 65 mol e$^-$ [mol C]$^{-1}$) – a reduction of approximately 30% (Fig. 7a, Supplementary Fig. S6). Similarly, corrected NPQ$_{NSV}$ also exhibited a narrower range (0.49 – 5.96) compared to pre-correction (0.49 – 8.7) (Fig 7b). Interestingly, when corrected for baseline fluorescence, the ability to predict $\Phi_{e,C}$ from all available variables was reduced across the whole dataset (from 54% to 46%) (Table 3), yet this was largely driven by data within cluster B where the best model from all variables explained only 34% of $\Phi_{e,C}$ variability after correction (Table 3). Notably, the corrected data still did not show any obvious trends with cell size (data not shown). Importantly, NPQ$_{NSV}$ remained the best overall predictor of $\Phi_{e,C}$ in this dataset even after $F_b$-correction, underscoring its potential value for retrieval of $\Phi_{e,C}$ during future field campaigns.

## 4. Discussion

Field-based campaigns have increasingly demonstrated that variability of $\Phi_{e,C}$ (the electron requirement for C-fixation) amongst natural phytoplankton communities can be explained by variance in prevailing environmental conditions (e.g. Lawrenz et al. 2013; Zhu et al. 2016, 2017, 2019), or more recently by photophysiological parameters ($NPQ_{NSV}$; Schuback et al. 2015) and traits governing resource acquisition (e.g. predominant cell size; Zhu et al. 2017). In this study, we demonstrated that $NPQ_{NSV}$ was a far improved predictor of $\Phi_{e,C}$ variability than prevailing environmental conditions across the physically complex Tasman Sea and EAC systems, determined by a high-throughput assessment of predominately surface waters. Specifically, $NPQ_{NSV}$ explained ~55% of observed variation across the dataset versus ~25% for environmental variables measured. $NPQ_{NSV}$ accounts for changes in both long-term driven acclimation in non-radiative decay as well as quasi-instantaneous PSII downregulation (see McKew et al. 2013), the latter of which is also highly-dependent upon the environmental history of the cells (Dimier et al. 2007; Queval and Foyer, 2012; Giovagnetti et al. 2014). As such, it is perhaps unsurprising that $NPQ_{NSV}$ ultimately proved a better predictor of $\Phi_{e,C}$ in this study, since the environmental descriptors used here typically account for prevailing, and not historical, environmental conditions. We did however attempt to account for previous light history of the sample by incubating as close as possible to the measured light-saturation parameter ($E_K$) – which can be interpreted as a simple indicator of photoacclimational status (Sakshaug et al. 1997) – and indeed, inclusion of $E/E_K$ as a predictor variable yielded no improvement in ability to predict $\Phi_{e,C}$. We further examined whether historic environmental conditions potentially influenced $\Phi_{e,C}$, by attempting to use remotely sensed data (temperature, light and Chl-$a$) to hind cast historic light and nutrient availability (not shown). However, this analysis was based on data restricted to certain time bins and ultimately failed to improve models, presumably due to the fact that such an approach did not take into account the position of phytoplankton cells over time and hence did not accurately capture the complex light and nutritional history experienced by cells in the water bodies sampled. Despite this, teasing apart prevailing versus historical conditioning of cell physiology (and its influence on $\Phi_{e,C}$) warrants more targeted investigation.

Although we included surface measurements of PAR at the time of sampling, our DistLM models did not include the diffuse attenuation coefficient for downwelling irradiance, $K_d$(PAR), which was not measured in this study, as a predictor variable. Whether inclusion of $K_d$(PAR) would have improved predictive retrieval of $\Phi_{e,C}$ is however uncertain, since this parameter was rarely identified to contribute significantly to empirical models developed by Lawrenz et al. (2013) during their global synthesis of $\Phi_{e,C}$ datasets, although was shown to be important in shelf waters and in the Baltic Sea (Lawrenz et al. 2013). Nevertheless, light availability has been shown to be an effective predictor of $\Phi_{e,C}$ over longer time scales of C-assimilation (e.g. 24 hr; see Zhu et al. 2016, 2017) and hence more prolonged environmental conditions driving acclimation states between phytoplankton communities along complex environmental gradients (Moore et al. 2003, 2006). As such, omission of $K_d$[PAR] and empirical descriptors of previous light history experienced by cells may have reduced the percentage of $\Phi_{e,C}$ variability explained by prevailing environmental conditions according to the models used in our study.

In the meta-analysis by Lawrenz et al. (2013), the greatest proportion of $\Phi_{e,C}$ variance explained by prevailing conditions was ~70%, and thus at face value, far higher than achieved with the current study. However, this appeared to be the exception rather than the rule in their meta-analysis, with many models actually performing far worse (either not statistically significant or explaining as little as 3% variability in $\Phi_{e,C}$). In fact, when averaged across all regions, prevailing conditions explained ~25% of $\Phi_{e,C}$ variance in their meta-analysis – relatively similar to findings in the present study. In contrast, recent studies (Schuback et al. 2015, 2016; Hughes et al. 2018b; Wei et al. 2019) have reported that 50-95% of $\Phi_{e,C}$ variance can be explained from knowledge of $NPQ_{NSV}$, thus collectively supporting the premise that the parameter $NPQ_{NSV}$ may be a promising predictor of $\Phi_{e,C}$ variability in the field. That being said, there are several important caveats which must be highlighted before reaching this conclusion:

Firstly, the calculations reported by Schuback et al. (2015, 2016; Schuback and Tortell, 2019) conflate variability of $\Phi_{e,C}$ and variability of PSII reaction centre content ($n_{PSII}$), whereas the present study inherently accounts for $n_{PSII}$ variability via fluorometric estimation of [RCII] via the Oxborough et al. (2012) algorithm. Schuback et al. (2015, 2016) convincingly demonstrate that $n_{PSII}$ is unlikely to be contributing to variance in their data through the calculation of "relative" PSU size, yet the same does not hold true for our study (see also Hughes et al. 2018b). In fact, when $n_{PSII}$ was removed from our calculations (thus $\Phi_{e,C}$ becomes $\Phi_{e,C}/n_{PSII}$), the predictive power of $NPQ_{NSV}$ weakened by two-fold for pooled data in this study ($r^2 = 0.21$, $p <0.01$; Fig. 6b). This suggests that fluorometric estimation of [RCII] is likely required across a more dynamic system such as the one examined in this study to effectively utilise $NPQ_{NSV}$ as a predictor of $\Phi_{e,C}$.

Secondly, as calculated values of both $NPQ_{NSV}$ and $vETR_{PSII}$ (and by extension $\Phi_{e,C}$, as: $vETR_{PSII}$/C-incorporation) both rely on the fluorescence parameter $F_v^{(')}$ (calculated as $F_m^{(')} - F_0^{(')}$) they are not strictly independent from one another. Moreover, $NPQ_{NSV}$ and $vETR_{PSII}$ as calculated via the absorption method of Oxborough et al. (2012), are sensitive to the presence of baseline fluorescence (see Oxborough et al. 2012; Boatman et al. 2019). Specifically, baseline fluorescence contribution results in an increase to measured $F_0^{(')}$ and $F_m^{(')}$ (thus a lower $F_v^{(')}/F_m^{(')}$), and artificially inflates both $NPQ_{NSV}$ and $vETR_{PSII}$ (Oxborough et al. 2012; Boatman et al. 2019). Incidentally, $vETR_{PSII}$ calculated according to the "Sigma method" (Kolber et al. 1998) is also not immune to effects of baseline fluorescence if a fluorometric estimation of [RCII] is included (Oxborough et al. 2012).

Baseline fluorescence is defined by Oxborough et al. (2012) as that emitted from sources other than functional PSII units, and may include: i) light-harvesting complexes separated from functional PSII units, ii) photoinactivated PSII units (Murphy et al. 2017), ii) Pigments outside of PSII (Campbell and Tyystjärvi, 2012) and iv) fluorescent dissolved organic matter (Cullen and Davies, 2003). In the present study, any contribution from fluorescent dissolved organic material was removed through subtraction of 0.2-µm filtered "blanks" which were subtracted from all samples according to routine FRRf protocol (see Hughes et al. 2018a). This procedure does not, however, remove potential contribution from baseline fluorescence of cellular origin (i.e. from uncoupled LHIIs or photoinactivated PSII units).

Dark-acclimated $F_v/F_m$ measurements of natural phytoplankton assemblages appear to be suppressed under nutrient-stress due to PSII photoinactivation (e.g. Moore et al. 2008) thus indicating that baseline fluorescence is a potential concern for field studies of $\Phi_{e,C}$. Even so, more recent work has explained observed empirical relationships between $NPQ_{NSV}$ and $vETR_{PSII}$ by the positive feedback link that exists between non-carbon fixing electron pathways and the build-up of ΔpH which activates certain components of NPQ (Nawrocki et al. 2015). In contrast, the role of baseline fluorescence in explaining such trends has been comparatively overlooked (but see Boatman et al. 2019). Interestingly, $NPQ_{NSV}$ is not a reliable predictor of $\Phi_{e,C}$ for unialgal phytoplankton strains grown under nutrient-replete, steady-state growth where the influence of baseline fluorescence is presumably minimal (see Hughes, 2018). However, it remains unclear whether this reflects limited influence of baseline fluorescence, or the fact that $NPQ_{NSV}$ is a good predictor of $\Phi_{e,C}$ only when one or more environmental stressors are at play, or indeed whether the relationship between $NPQ_{NSV}$ and $\Phi_{e,C}$ breaks down under the influence of a dominant species-specific $NPQ_{NSV}$ signature.

Thirdly, and perhaps most importantly, the slopes of the relationships between $\Phi_{e,C}$ and NPQ differ considerably between studies (Schuback et al. 2015, 2016; Hughes et al. 2018b), likely as a result of one or more of the aforementioned factors. Critically, this limits the applicability of NPQ$_{NSV}$ as a viable standalone predictor of $\Phi_{e,C}$, unless we can better understand and predict the slope of this relationship at any given point in space and time.

*4.1. Correcting for baseline fluorescence (and package effect)*

Our sensitivity analysis demonstrated how baseline fluorescence ($F_b$) could potentially contribute to strong correlations between NPQ$_{NSV}$ and $\Phi_{e,C}$ observed in recent field studies (e.g. Hughes et al. 2018b), by artificially forcing an increase in both values, which critically, may also yield inflated estimates of $\Phi_{e,C}$. This is a potentially significant finding depending on whether FRRf users wish to i) simply convert measured vETR$_{PSII}$ to a measure of C-fixation via a simple conversion factor or ii) gain mechanistic insight into the factors regulating "true" $\Phi_{e,C}$ – i.e. the absolute number of electrons invested into carbon biomass. For the former, an artificially-inflated measured $\Phi_{e,C}$ may ultimately exhibit the strongest correlation with NPQ$_{NSV}$, although this approach ultimately hinges on being able to reliably predict the slope of this relationship at any given point in space and time. For the latter, $F_b$-correction (and indeed correction for packaging effects, discussed below) will likely prove critical steps that become routinely incorporation into existing FRRf "best practice" recommendations (e.g. Hughes et al. 2018a). Of course, this would likely need to occur over a prolonged timeframe, depending on a number of factors not limited to: the life cycle and upgradability of existing instrumentation, generational evolution of single-turnover active fluorometers and indeed wider evaluation of the corrective procedures outlined in Boatman et al. (2019) in studies of natural phytoplankton assemblages. Certainly, future studies should fundamentally report the range of both corrected and non-corrected data (as we have done here) to advance our understanding of the magnitude and extent to which $F_b$ (and the package effect) may influence FRRf-datasets.

In addition to correcting for baseline fluorescence influence, Boatman et al. (2019) also demonstrated that cellular packaging effects contribute to errors in $v$ETR$_{PSII}$ determined by the absorption method of Oxborough et al (2012). Specifically, the reabsorption of PSII fluorescence by pigments within the cell modifies the relationship between PSII photochemistry and fluorescence emission, thus violating a key assumption of the absorption algorithm (see Oxborough et al. 2012). Unfortunately, it was not possible to retrospectively apply the corrective procedure suggested by Boatman et al. (2019) to this dataset as it requires measurements from a fluorometer fitted with narrow bandpass filters (680 and 730nm). Clearly this warrants further examination for future field-based studies spanning broad environmental conditions and hence phytoplankton cell size distributions.

*4.2. Cell size does not appear to significantly aid retrieval of $\Phi_{e,C}$*

Phytoplankton cell size varies by 8 orders of magnitude, and governs a number of physiological characteristics including photosynthetic performance (Sarthou et al. 2005; Finkel et al. 2009). Many traits potentially linked with photosynthetic performance including: PSII absorption efficiency, PSII efficiency and nutrient uptake rate, have been shown to scale allometrically with cell volume (Ciotti et al. 2002; Suggett et al. 2009b, 2015; Litchman et al. 2007). Consequently, photosynthetic rates normalised to cellular volume are often lower for larger phytoplankton (Bouman et al. 2005; Barnes et al. 2015) due to biophysical constraints upon light absorption and nutrient-uptake, through reduced surface-area-to-volume ratios (Marra et al. 2007). However, whilst trends in $\Phi_{e,C}$ were apparent when data were binned according to dominant Chl-*a* size fraction in this study, overall we did not establish a strong relationship between $\Phi_{e,C}$ and cell size, thus retrieval of $\Phi_{e,C}$ was only marginally improved with a descriptor of cell size, similar to findings of Zhu et al. (2017).

We utilised size-fractionated Chl-*a* as a routinely measured descriptor of phytoplankton community size structure during our study. Aside from the disadvantage of information loss regarding species composition, the size-fractionated Chl-*a* method relies on the assumption that Chl-*a* biomass is directly related to primary productivity. Yet, reliability of Chl-*a* as a pigment indicator for carbon fixation rates has been questioned (Bassett, 2015). In examining data from 27 studies, Bassett (2015) demonstrated that a strong relationship between Chl-*a* biomass and $^{14}C$ production was evident for just 17% of the data, suggesting a large contribution of a given size fraction does not necessarily translate to an equivalent proportion of total production, and is likely strongly dependent upon physical and biological conditions (see also Pommier et al. 2008).

*4.3. Using $\Phi_{e,C}$ to inform dynamics of the study region*

In contrast to previous FRRf-based $\Phi_{e,C}$ field studies (e.g. Robinson et al. 2014; Zhu et al. 2017), we did not observe measurements of $\Phi_{e,C}$ to fall below the theoretical minimum of 4 mol e$^-$ (mol C)$^{-1}$, presumably since methodological contributors were kept to a minimum through: (i) use of coupled FRRf-$^{14}C$ incubations, (ii) application of spectral correction factors, and (iii) avoiding using assumed values of n$_{PSII}$ (discussed by Robinson et al. 2009; see also Suggett et al. 2004). Previous laboratory FRRf-based $\Phi_{e,C}$ studies that have also employed a "dual incubation" approach and therefore also removed such potential sources of error (Suggett et al. 2009a) similarly measured few values of $\Phi_{e,C}$ <4 (<5% of *n* = 48). The wide range of $\Phi_{e,C}$ values measured in our study (~4 - 65 mol e$^-$ [mol C]$^{-1}$) likely reflects the well-documented spatial and temporal complexity of physico-chemical conditions within the study area (Baird et al. 2011; Hassler et al. 2011), and is consistent with other datasets that also span broad changes in environmental condition (e.g. Lawrenz et al. 2013; Zhu et al. 2017; Ko et al. 2019). Observed patterns in photophysiology however, were not as clearly evident as those for physico-chemical variables across the study area (Figs 2a-f, 3a-f), and we later discuss a possible role for diurnal effects on measured photophysiological parameters.

Initial analysis of the prevailing environmental conditions identified two discrete environmental regimes. Specifically, these were characterised by warmer, low-nutrient water dominated by small cells versus cooler, nutrient-rich waters dominated by larger cells: our clusters A, and B respectively. In turn, these were characterised by communities exhibiting a higher $\Phi_{e,C}$ in the nutrient-poor, small cell-dominated cluster A, thus appearing to support broader observations of taxonomy with $\Phi_{e,C}$ (Robinson et al. 2014; Zhu et al. 2017). This is also consistent with the oceanography of the region, whereby the EAC is warm, relatively low in nutrients and dominated by smaller cells (Hassler et al. 2011), whilst the Tasman Sea water mass is characterised by lower temperatures, increased nutrient availability, and a phytoplankton community often dominated by larger cells (Baird et al. 2008). Such an outcome is entirely consistent with previous work that has binned data into water masses (Lawrenz et al. 2013; Robinson et al. 2014) or environmental regime (Zhu et al. 2017). However, it also consistent with these prior studies that have also shown environmental condition alone to be a poor predictor of $\Phi_{e,C}$ (mean $r^2$ = ~25%). Thus, $\Phi_{e,C}$ appears to follow broad oceanographic trends with the prevailing environmental conditions measured, but ultimately requires a physiological descriptor of both prevailing (short-term dynamic physiological regulation) and historical conditions that drive acclimation states, such as $NPQ_{NSV}$. This presumably explains why cell size alone may also fail to provide a robust descriptor of $\Phi_{e,C}$ (see Zhu et al. 2017).

*4.4. Consideration of incubation conditions, carbon lifetimes & diurnal effects*

While our reported values of $\Phi_{e,C}$ correspond well with previous FRRf field studies (e.g. Lawrenz et al. 2013; Zhu et al. 2016; Ko et al. 2019) it is important to consider key differences in incubation conditions in the present study. Due to the "dual incubation" approach used we incubated samples under a single irradiance - corresponding to the light saturation parameter ($E_K$) - rather than deriving $\Phi_{e,C}$ from photosynthesis-irradiance (PE) curves. Indeed, largest divergence of $\Phi_{e,C}$ from the theoretical minimum ratio of 4 mol e$^-$ (mol $CO_2$)$^{-1}$ has most commonly been observed under saturating light (Schuback et al. 2016; Zhu et al. 2016), so it is notable that we observed significant decoupling of electrons and carbon incubating at $E_K$. Although a range of environmental factors (e.g. nitrogen limitation; Hughes et al. 2018b), and indeed taxonomy also appear to drive increases in $\Phi_{e,C}$ (reviewed by Hughes et al. 2018a), an obvious avenue of exploration to explain high $\Phi_{e,C}$ value in this study is influence of methodology – in particular surrounding quantification of C-fixation rates.

A historical challenge for oceanographers using the $^{14}$C-method is the complex relationship between incubation length and cellular retention time of fixed-C that can introduce uncertainty in $\Phi_{e,C}$ measurements (Hughes et al. 2018a). Variability in phytoplankton growth rates influences the lifetime of newly-fixed carbon (Halsey et al. 2010, 2011, 2013), and the $^{14}$C-method may measure a C-fixation rate somewhere between gross and net carbon production, unless either very short (<20 min) or long (>12 hr) incubations are used to reliably target either process respectively. Studies examining $\Phi_{e,C}$ in the field to date have routinely intermediate incubation lengths (1-4 hr) (Lawrenz et al. 2013), necessitated by low phytoplankton biomass, small sample size or limited $^{14}$C activity per sample due to radioisotope handling protocols on research vessels (Hughes et al. 2018a). An incubation time of two hours was used in the present study due to a combination of aforementioned factors, thus variation in carbon lifetimes may introduce a degree of variability in measured $\Phi_{e,C}$ between samples. Previous experimental work using natural phytoplankton communities from the Australian coast found an incubation length of two hours routinely resulted in a measured $^{14}$C-fixation rate close to net, rather than gross, carbon production (Hughes et al. 2018a). If that holds true for this dataset too, uncertainty in the current study would be expected to be relatively small, yet the comparison of ETR against net carbon production would result in overall inflated measurements of $\Phi_{e,C}$.

Recent FRRf studies have also highlighted the importance to consider diurnal effects on photophysiology (e.g. Aardema et al. 2019), as many measured photophysiological paramaters exhibit diurnal trends, including $\Phi_{e,C}$ (Schuback et al. 2016; Aardema et al. 2019), linked to circadian rhythms, cell cycle status and short-term photoacclimatory responses (Behrenfeld et al. 2002; Cohen and Golden, 2015; Schuback et al. 2016). Schuback et al. (2016) convincingly demonstrated that $NPQ_{NSV}$ was a good predictor of $\Phi_{e,C}$ over a diurnal sampling period, and thus, provides an effective means to examine spatial variability in $\Phi_{e,C}$ – an approach that we have employed here. Nevertheless, the robustness of $NPQ_{NSV}$ as a means to assess spatial variability of $\Phi_{e,C}$ requires further targeted examination (see Aardema et al. 2019), and therefore it is entirely plausible that diurnal effects may contribute to variability in $\Phi_{e,C}$ reported in this study (see Supplementary Fig. S7).

*4.5. Conclusions*

Using a unique high-throughput "dual incubation" FRRf approach to retrieve values of $\Phi_{e,C}$, we have demonstrated co-variance of $\Phi_{e,C}$ with independently measured environmental variables (notably temperature, salinity and $PO_4^{3-}$), but particularly with the physiologically-dependent variable ($NPQ_{NSV}$). The latter outcome suggests that for our dataset, a modification of Eq. 3 with the calculation of $NPQ_{NSV}$ ($F_0´=F_v´$) (McKew et al. 2013) adjusted for the specific relationship describing dependency between $\Phi_{e,C}$ and $NPQ_{NSV}$ (Fig. 6a) could redefine the FRRf algorithm to retrieve C-fixation rates within this study region, yet this clearly warrants further validation. That said, we further demonstrate that baseline fluorescence can artificially strengthen the relationship between $\Phi_{e,C}$ and $NPQ_{NSV}$ which may prove advantageous for empirical prediction of C-fixation rates from FRRf datasets yet also lead to inflated estimates of $\Phi_{e,C}$ unless correction procedures are applied (Boatman et al. 2019). A logical next step would therefore be the retrospective incorporation of both $NPQ_{NSV}$ measurements and corrective procedures outlined above to historic FRRf–C-uptake data campaigns (e.g. those collated by Lawrenz et al. 2013). This will allow for additional comparison between the respective performance of $NPQ_{NSV}$ and prevailing conditions in predicting $\Phi_{e,C}$ variability across marine provinces (for both corrected and non-corrected FRRf data). Such steps will permit critical examination of the apparent discrepancy in the slopes of the relationship between $\Phi_{e,C}$ and $NPQ_{NSV}$ (see Schuback et al. 2015; Hughes et al. 2018b) and hence 'global' applicability of this approach through autonomous FRRf deployments.

**Acknowledgements**

The authors would like to thank the crew of the RV Investigator and the Marine National Facility (MNF) for their support during the sampling campaign. We also wish to acknowledge invaluable assistance from Justin Ashworth regarding onboard phytoplankton identification. Marco Alvarez and Marco Giardina for providing technical support and advice at various stages during the research voyage. We also acknowledge valuable discussions across the peer community in distilling the concepts presented here; in particular, we thank Doug Campbell and Mark Moore. Contribution of DJS was supported by an ARC Future Fellowship (FT130100202), and input of MD and DJS enhanced through involvement with an ARC Linkage Infrastructure, Equipment and Facilities project LE160100146 led by David Antoine.

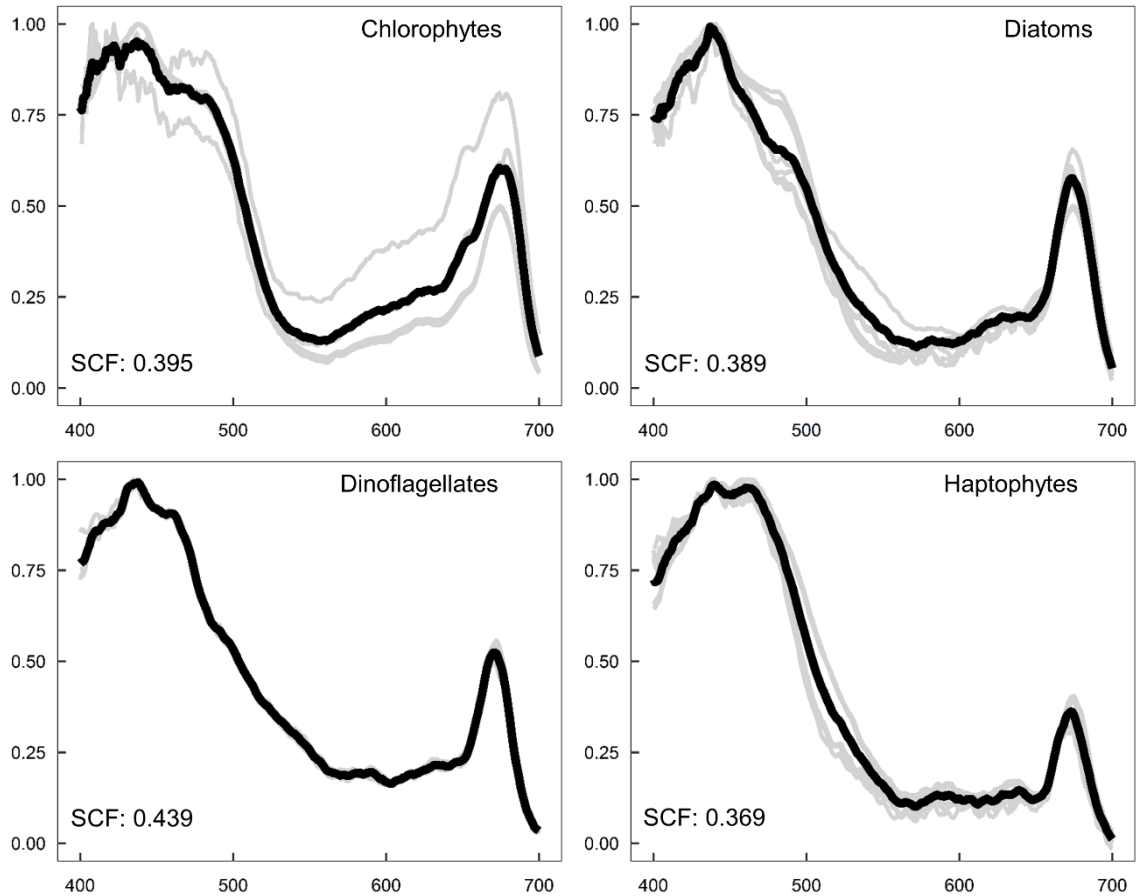

**Supplementary Figure S1** Phytoplankton fluorescence excitation spectra measured from phytoplankton cultures used to derive spectral correction factors (SCFs) for dominant phytoplankton groups observed during the main study. Fluorescence excitation spectra (400 – 700 nm) were measured after treatment of with 3-(3,4-dichlorophenyl)-1,1-dimethylurea (DCMU) for several representatives of each taxonomic group. Light grey lines denote measurements for each species within the group ($n = 4$) with the exception of dinoflagellates ($n = 2$), while black line denotes the mean excitation spectra (error not shown for clarity) that was ultimately used to calculate the SCF for that taxonomic group (bottom left corner). SCFs were used to spectrally-adjust values of the absorption coefficient for PSII light harvesting, $a_{LHII}$ prior to calculation of electron transport rates (vETR$_{PSII}$).

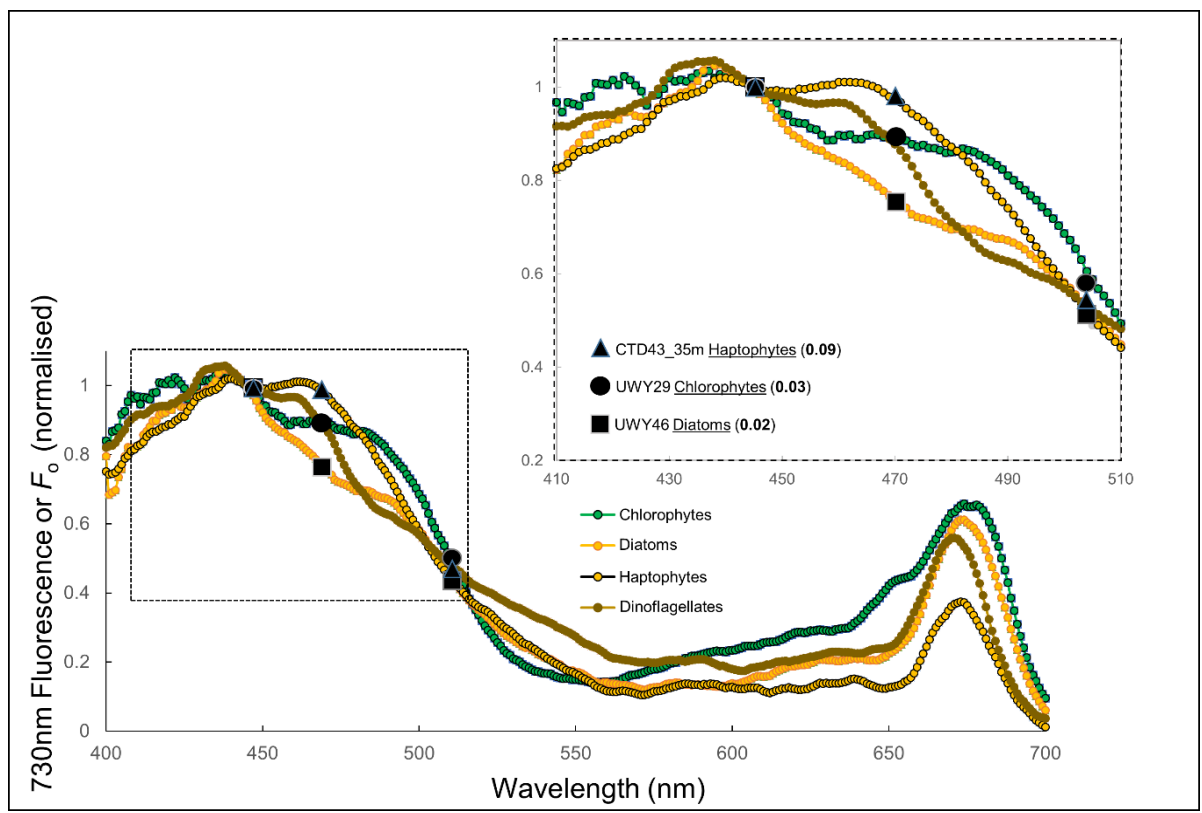

**Supplementary Figure S2**. Fluorescence excitation spectra (normalised to 445 nm) for the four dominant phytoplankton groups observed in this study (chlorophytes, diatoms, haptophytes and dinoflagellates) from which spectral correction factors (SCFs) were derived. For samples where it was not possible to ascertain phytoplankton dominant groups by microscopy ($n = 11$), the relevant SCF was selected based on how closely minimal fluorescence ($F_o$) at 445, 470 and 505 nm matched the spectral fluorescence for each representative excitation spectra (when both were normalized to 445 nm). $F_o$ was measured with Fast Repetition Rate fluorometer (Soliense Benchtop Marine Lift-FRR) for three representative samples, CTD43_35m (black triangles), UWY29 (black circles) and UWY46 (black squares) are shown overlaid on the fluorescence excitation spectra (see break-out panel). The dominant taxa (underlined text) for each 11 samples was selected based on whichever taxonomic group exhibited the least cumulative difference between fluorescence at 445, 470 and 505 nm wavelengths (differences are shown in brackets for the three samples displayed). In total, observed cumulative differences for identified dominant taxonomic group ranged from 0.02 – 0.15 (mean: 0.08). Once the dominant phytoplankton group was identified, the relevant SCF was applied to adjust electron transport rates ($\nu ETR_{PSII}$) (see Supplementary Fig S1).

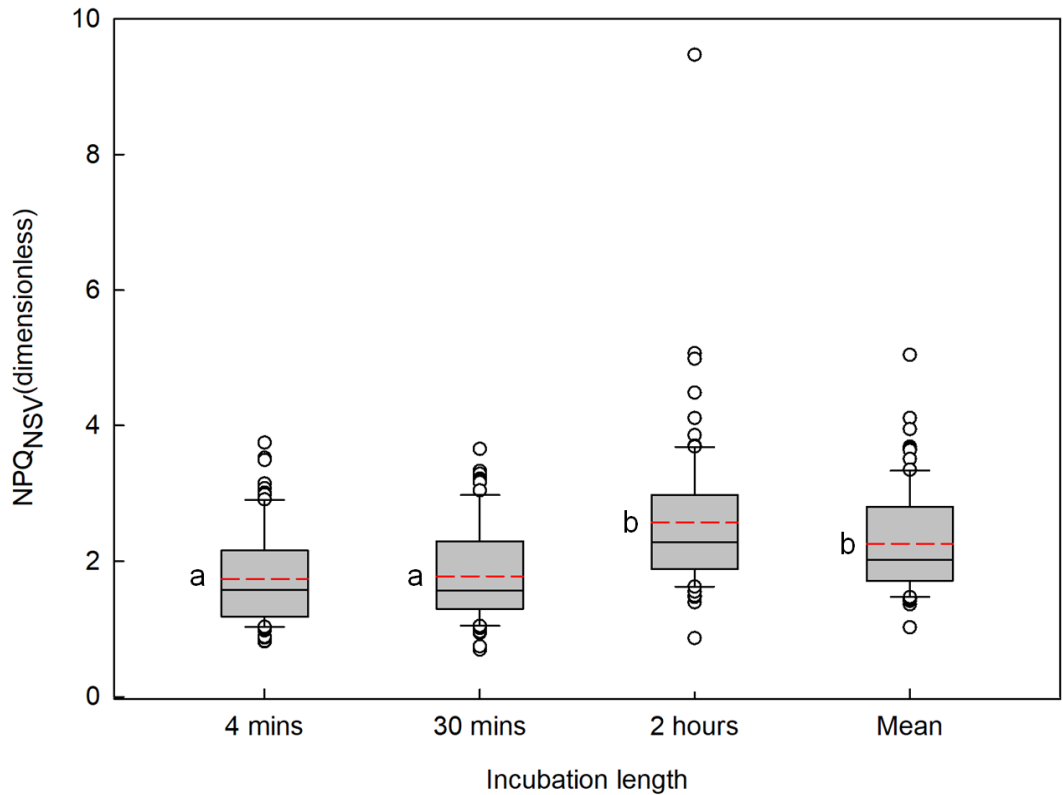

**Supplementary Figure S3**. Boxplot of measured normalised Stern-Volmer quenching (NPQ$_{NSV}$) during a two hour incubation for all samples in this study ($n = 80$) at intervals of 4 min, 30 min and 120 min – where values are averaged from the last three acquisitions at that point in time). Also shown is the average (mean) value over the entire incubation for all samples. The length of the box corresponds to the interquartile range, whiskers represent the 10[th] and 90[th] percentile, solid black line denotes the median value, red dashed line shows the mean and open circles indicate outliers. Letters alongside boxes indicate means that are statistically indistinguishable (Kruskal-Wallis test, Dunn's multiple comparison test, α=0.05).

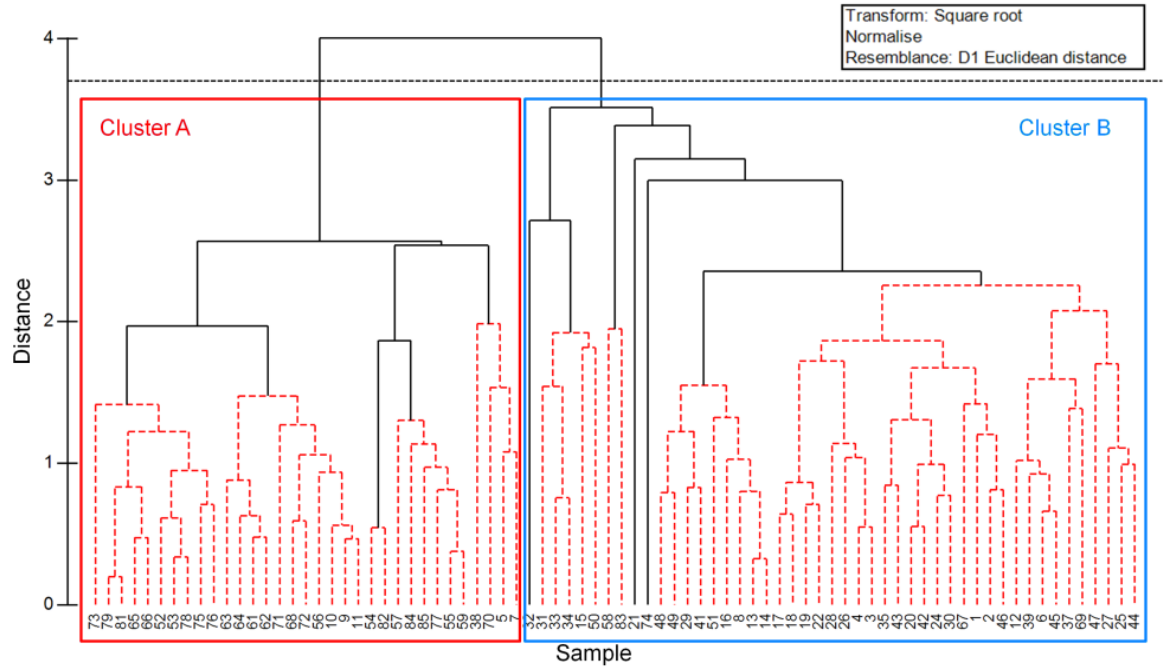

**Supplementary Figure S4** Hierarchical cluster analysis (HCA) of samples ($n = 80$) based upon physic-chemical variables (temperature, salinity, $NH_4^+$, $NO_x^-$, $PO_4^{3-}$ and Si). HCA identified two distinct hydrographic clusters, **a** and **b** (highlighted by blue and green sections respectively).

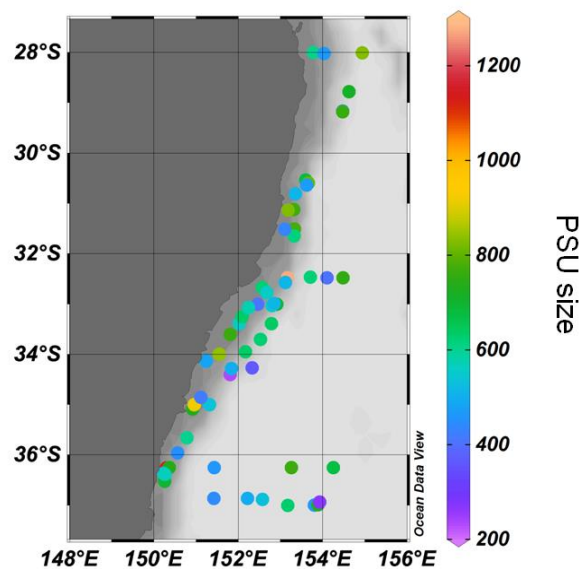

**Supplementary Figure S5** Photosynthetic unit (PSU) size surface water sampled in coastal, Eastern Australian Current (EAC) and Tasman Sea water masses measured from the RV *Investigator* (August – September 2016, IN2016_v03

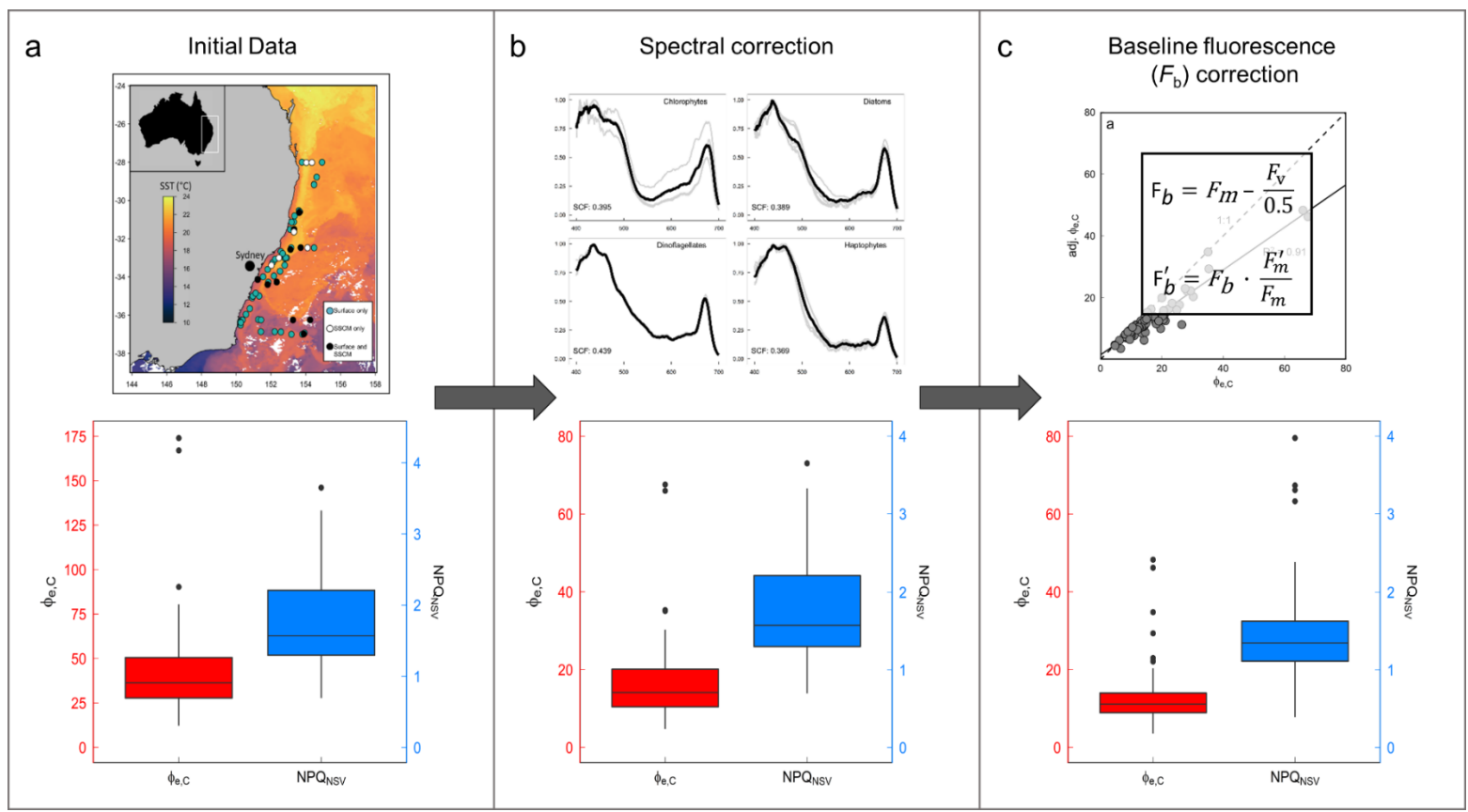

**Supplementary Figure S6** Dual-axis boxplots showing measured electron requirement for carbon fixation ($\Phi_{e,C}$) and normalised Stern-Volmer quenching ($NPQ_{NSV}$) a) prior to spectral correction, b) after application of spectral correction factors (S.C.F) to adjust electron transport rates (and thus $\Phi_{e,C}$) according to the dominant phytoplankton taxa present in the sample and c) after further correction for contribution of baseline fluorescence as proposed by Boatman et al. (2019). Note the different scale for the Y-axis in panel a ($\Phi_{e,C}$).

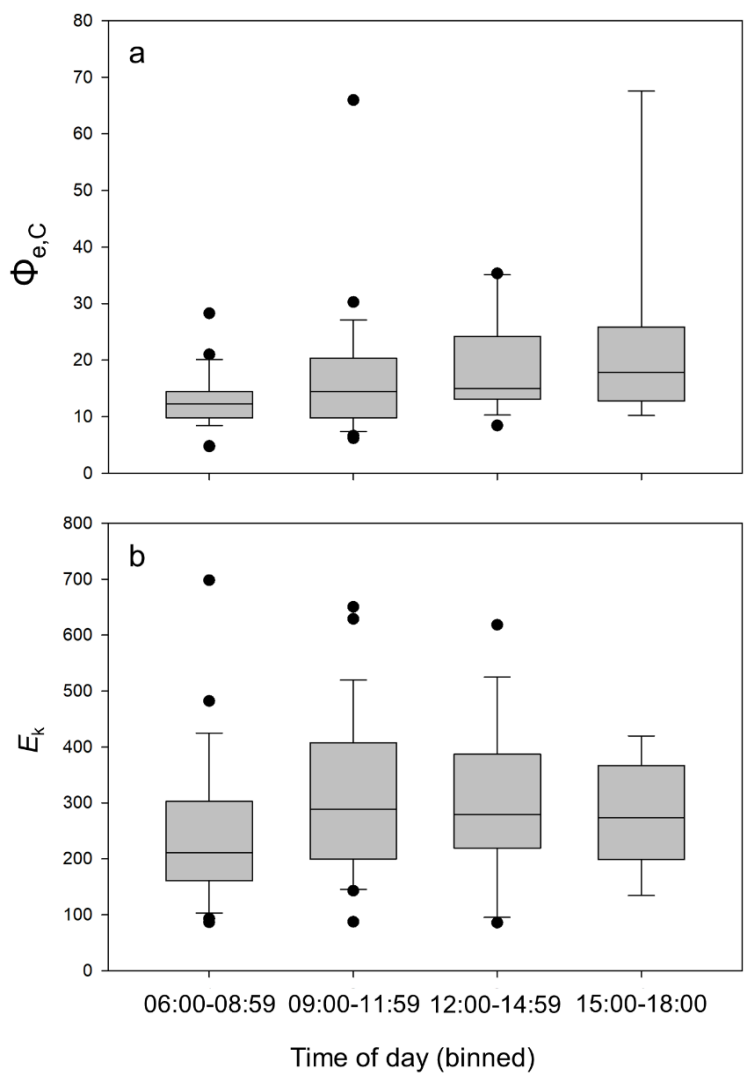

**Supplementary Figure S7** a) the electron requirement for carbon fixation ($\Phi_{e,C}$) and b) light saturation parameter ($E_K$) for all data in this study binned into time periods, early morning (06:00–08:59, $n=28$), late morning (09:00-11:59, $n=26$), early afternoon (12:00-14:59, $n=18$) and late afternoon (15:00-18:00, $n=8$). A Kruskal-Wallis test detected significant differences between for groups $\Phi_{e,C}$ ($p <0.05$)(although post-hoc tests were unable to resolve specific differences between time bins). No significant differences were detected for $E_K$.